\documentclass[12pt]{article}
\usepackage[utf8]{inputenc}
\usepackage{hyperref}
\usepackage{cite}
\usepackage{verbatim,graphics,graphicx,color,slashed,amsmath}
\usepackage[T1]{fontenc}
\usepackage{subfigure}
\usepackage[table,svgnames]{xcolor}
\usepackage{color}
\usepackage{amsfonts}
\usepackage{amssymb}
\usepackage{bbding}
\usepackage{appendix}
\usepackage{mhchem}
\usepackage{stmaryrd}
\usepackage{blkarray}
\usepackage[export]{adjustbox}
\graphicspath{ {./images/} }
\usepackage{bbold}
\usepackage{CJKutf8}
\usepackage[T1]{fontenc}
\usepackage{bm}

\allowdisplaybreaks[1]

\def\thefootnote{\fnsymbol{footnote}}

\usepackage{multirow}

\newcommand{\be}{\begin{equation}}
\newcommand{\ee}{\end{equation}}
\newcommand{\bea}{\begin{eqnarray}}
\newcommand{\eea}{\end{eqnarray}}

\usepackage[normalem]{ulem}

\newcommand{\tr}{\text{tr}}

\newcommand{\rmI}{{\rm i}}

\newcommand{\mc}{\mathcal}
\newcommand{\bl}{\mathbf}

\begin{document}

\begin{center}
{\Large\bf One-loop Matching of the Type-III Seesaw Model onto \\the Standard Model Effective Field Theory}
\end{center}

\vspace{0.2cm}

\begin{center}
{\bf Xu Li }~\footnote{E-mail: lixu96@ihep.ac.cn},
\quad
{\bf Shun Zhou}~\footnote{E-mail: zhoush@ihep.ac.cn}
\\
\vspace{0.2cm}
{\small
Institute of High Energy Physics, Chinese Academy of Sciences, Beijing 100049, China\\
School of Physical Sciences, University of Chinese Academy of Sciences, Beijing 100049, China}
\end{center}

\vspace{1.5cm}

\begin{abstract}
In previous works~\cite{Zhang:2021jdf, Li:2022ipc}, we have performed the one-loop matching of both type-I and type-II seesaw models for neutrino masses onto the Standard Model Effective Field Theories (SMEFT). In the present paper, by matching the type-III seesaw model onto the SMEFT at the one-loop level, we complete this series of studies on the construction of low-energy effective field theories (EFTs) for the canonical seesaw models. After integrating out the heavy fermionic triplets in the type-III seesaw model via both functional and diagrammatic approaches, we find 33 dimension-six (dim-6) operators in the Warsaw basis and their Wilson coefficients, while the number of dim-6 operators is 31 (or 41) for the EFT of type-I (or type-II) seesaw model. Furthermore, we calculate the branching ratios of radiative decays of charged leptons in the EFT. Then, the relationship between the beta function of the quartic Higgs coupling $\lambda$ in the full theory and that of $\lambda^{}_{\rm EFT}$ in the EFT is clarified. Finally, we briefly discuss the phenomenological implications of three types of seesaw EFTs and propose working observables that are sensitive to the four-fermion operators, which could be used to distinguish among different seesaw models in collider experiments.
\end{abstract}


\def\thefootnote{\arabic{footnote}}
\setcounter{footnote}{0}
\newpage

\section{Introduction}
\label{sec1}
In the past few decades, a large number of neutrino oscillation experiments have provided us with strong evidence that neutrinos are massive but extremely light, and the lepton flavor mixing is very significant~\cite{Xing:2020ijf}. To accommodate tiny neutrino masses, three types of canonical seesaw models have been proposed~\cite{Minkowski:1977sc, Yanagida:1979as, Gell-Mann:1979vob, Glashow:1979nm, Mohapatra:1979ia,Konetschny:1977bn, Magg:1980ut, Schechter:1980gr, Cheng:1980qt, Mohapatra:1980yp, Lazarides:1980nt,Foot:1988aq}. In these seesaw models, neutrinos turn out to be Majorana particles and the lightness of their masses can be ascribed to the largeness of heavy particle masses, which may be far beyond the electroweak scale $\Lambda^{}_{\rm EW}\sim 100~{\rm GeV}$. In this case, it seems quite difficult to achieve the on-shell production of these heavy particles in current or even next-generation collider experiments. An alternative way to scrutinize the seesaw models of neutrino masses is to explore the off-shell effects of heavy states, which may lead to remarkable deviations of some observables from the predictions of the Standard Model (SM).

At this point, the SM effective field theory (SMEFT) is a well-established theoretical framework to interpret possible deviations from the SM predictions~\cite{Buchmuller:1985jz,Brivio:2017vri, Isidori:2023pyp}. Respecting the SM gauge symmetries and particle content, the SMEFT supplements the SM with a series of higher-dimensional operators that are suppressed by the inverse power of the cutoff scale $\Lambda$. At the leading order of $1/\Lambda$, there is a unique dimension-five (dim-5) operator, i.e., the Weinberg operator~\cite{Weinberg:1979sa}, which gives rise to tiny neutrino masses after the spontaneous gauge symmetry breaking. The discovery of neutrino oscillations can be regarded as the robust evidence for the Weinberg operator. In the literature, a lot of efforts have been made to find the minimal complete set of operators of various mass dimensions in the SMEFT~\cite{Weinberg:1979sa,Grzadkowski:2010es,Lehman:2014jma, Li:2020gnx,Murphy:2020rsh,Liao:2020jmn,Li:2020xlh,Harlander:2023psl}.
Nevertheless, as the dimension of the operator becomes higher, the number of independent operators in the SMEFT will increase rapidly~\cite{Henning:2015alf}. Fortunately, such operators are also strongly suppressed by higher powers of $1/\Lambda$, rendering their impact on physical observables smaller. For this reason, the most commonly used basis is the Warsaw basis~\cite{Buchmuller:1985jz} of dim-6, including 59 independent baryon-number-conserving operators and 4 baryon-number-violating ones. In the near future, the experiments at the high-energy and high-intensity frontiers will hopefully measure the Wilson coefficients of these dim-6 operators and thus offer useful information about the ultraviolet (UV) full theory, from which the relevant dim-6 operators arise.

On the other hand, provided a well-motivated and renormalizable UV theory, one can match it onto the SMEFT by integrating out heavy degrees of freedom above the cutoff scale $\Lambda$. In this way, the resultant Wilson coefficients of higher-dimensional operators are completely determined by the model parameters in the UV theory and thus highly correlated. Ref.~\cite{Weinberg:1980wa} outlines the basic strategy to construct the EFT from a specific renormalizable UV theory. The tree-level contribution of any UV completion (with general scalar, spinor and vector fields and any types of interactions) to the Wilson coefficients of dim-6 operators in the SMEFT has been derived in Ref.~\cite{deBlas:2017xtg}. At the loop level, many UV models as the extensions of SM are found to be able to generate those dim-6 operators via one-loop matching~\cite{Henning:2014wua,Drozd:2015rsp,Ellis:2016enq,Henning:2016lyp,Fuentes-Martin:2016uol,Zhang:2016pja,DasBakshi:2018vni,Ellis:2017jns,Kramer:2019fwz,Cohen:2019btp,Cohen:2020fcu,Cohen:2020qvb,Fuentes-Martin:2020udw,Dittmaier:2021fls,Brivio:2021alv,Dedes:2021abc,Coy:2021hyr,Ohlsson:2022hfl,Du:2022vso,Zhang:2022osj,Liao:2022cwh}.

In view of the discovery of neutrino oscillations, the canonical seesaw models are indeed well-motivated UV theories. The one-loop matching of type-I and type-II seesaw models onto the SMEFT has been carried out in previous works~\cite{Zhang:2021jdf,Li:2022ipc}. The EFT descriptions of the seesaw models are referred to as the seesaw effective field theories (SEFTs).
In this paper, we continue to carry out the complete one-loop matching for the type-III seesaw model up to dim-6 effective operators. To this end, both functional and diagrammatic methods are implemented to integrate out the heavy fermionic triplets in the type-III seesaw model. As a cross-check, the results by using these two different methods are compared with each other. Several mistakes in Ref.~\cite{Du:2022vso}, where the one-loop matching of the type-III seesaw model has been done via the functional approach, are corrected. The main motivation for such an investigation is two-fold. First, the fermionic triplets in the type-III seesaw model experience the SM gauge interactions and come with three flavors, so the one-loop construction of its low-energy EFT is more complicated than that for either type-I or type-II seesaw models. Second, only with all these types of EFTs for the seesaw models can one start to consider whether it is possible to find out physical observables at low energies to distinguish among three types of seesaw models. Further discussions about the SEFT-III are also given. We work in the one-loop SEFT-III and calculate the rates of lepton-flavor-violating rare decays of charged leptons, such as $\mu \to e \gamma$. It is demonstrated that the SEFT-III can exactly reproduce the results in the full theory in the large-mass limit. Then, the beta function of the quartic Higgs coupling $\lambda^{}_{\rm EFT}$ is discussed and related to that in the full theory. By doing so, we show that the one-loop matching plays an important role in reconstructing the beta function in the full theory from that in the EFT. Finally, based on the dim-6 four-fermion operators, a brief study on the possible way to distinguish among three types of seesaw models is presented.

The remaining part of this paper is structured as follows. In Sec.~\ref{sec:model}, we introduce the type-III seesaw model and establish our notations and conventions. Then, the tree-level matching up to dim-6 is accomplished by utilizing the equation of motion. The one-loop matching onto the SMEFT with functional and diagrammatic approaches is given in Sec.~\ref{result}. The resulting dim-6 operators in the Warsaw basis and their Wilson coefficients are explicitly provided. In Sec.~\ref{sec:discuss}, we further discuss some phenomenological implications of the SEFTs. We summarize the main results and conclude in Sec.~\ref{sec:conclude}.

\section{The Type-III Seesaw Model}
\label{sec:model}
The type-III seesaw model extends the SM by introducing three right-handed fermionic triplets $\Sigma^{}_{\rm R}$ in the adjoint representation of the ${\rm SU}(2)^{}_{\rm L}$ gauge group and with the hypercharge $Y=0$. The gauge-invariant Lagrangian for the full type-III seesaw model is written as~\cite{Foot:1988aq}
\bea
\mc{L}^{}_{\rm type-III} &=& \mc{L}^{}_{\rm SM} + \text{Tr} \left({\overline{\mathrm{\Sigma}^{}_{\rm R}}} \rmI \slashed{D} \mathrm{\Sigma}^{}_{\rm R}\right) - \left[\sqrt2\ \overline{\ell^{}_{\rm L}}Y^{}_\mathrm{\Sigma} \mathrm{\Sigma}^{}_{\rm R} \widetilde{H} + \frac{1}{2} \text{Tr} \left({\overline{\mathrm{\Sigma}_{\rm R}^{\rm c}}}{M}^{}_\mathrm{\Sigma} \mathrm{\Sigma}^{}_{\rm R}\right) + \text{h.c.}\right]\;,
\label{eq:Luv}
\eea
with the SM Lagrangian being
\begin{eqnarray}
	\mathcal{L}^{}_{\rm SM} &=& -\frac{1}{4} G^A_{\mu\nu} G^{A\mu\nu} -\frac{1}{4} W^I_{\mu\nu} W^{I\mu\nu} -\frac{1}{4} B^{}_{\mu\nu} B^{\mu\nu} + \left( D^{}_\mu H \right)^\dagger \left( D^\mu H \right) - m^2H^\dag H - \lambda\left(H^\dag H\right)^2 \nonumber
	\\
	&&
	+ \sum^{}_f \overline{f} \rmI \slashed{D} f - \left( \overline{Q^{}_{\rm L}} Y^{}_{\rm u} \widetilde{H} U^{}_{\rm R} + \overline{Q^{}_{\rm L}} Y^{}_{\rm d} H D^{}_{\rm R} + \overline{\ell^{}_{\rm L}} Y^{}_l H E^{}_{\rm R} + {\rm h.c.} \right) \;,
	\label{eq:Lagrangian-SM}
\end{eqnarray}
where $f=Q^{}_{\rm L}, U^{}_{\rm R}, D^{}_{\rm R}, \ell^{}_{\rm L}, E^{}_{\rm R}$ refer to the SM fermionic doublets and singlets, and $H$ is the SM Higgs doublet with $Y = 1/2$ while $\widetilde{H}$ is defined as $\widetilde{H} \equiv \rmI \sigma^2 H^*$. In addition, $\ell^{\rm c}_{\rm L} \equiv {\sf C} \overline{\ell^{}_{\rm L}}^{\rm T}$ and $\Sigma^{\rm c}_{\rm R} \equiv {\sf C}\overline{\Sigma^{}_{\rm R}}^{\rm T}$ are defined, where ${\sf C}\equiv \rmI \gamma^2 \gamma^0$ denotes the charge-conjugation matrix. Without loss of generality, we shall work in the mass basis of $\Sigma_{\rm R}^{}$, namely, the mass matrix for $\Sigma_{\rm R}^{}$ is diagonal with ${M}^{}_\Sigma = \text{diag}\{M^{}_1, M^{}_2, M^{}_3\}$. It is worthwhile to note that three heavy states in the same fermionic triplet have the same mass, while three flavors of fermionic triplets have their own masses as indicated by their mass matrix.

In Eq.~\eqref{eq:Luv}, the covariant derivative reads $D^{}_\mu \equiv \partial^{}_\mu - \rmI g^{}_1 Y B^{}_\mu - \rmI g^{}_2 T^I W^I_\mu - \rmI g^{}_{\rm s} T^A G^A_\mu$ as usual. For the fields in the fundamental representation of ${\rm SU}(2)^{}_{\rm L}$, we have $T^I = \sigma^I/2$ (for $I = 1, 2, 3$) with $\sigma^I$ being the Pauli matrices. For the adjoint representation, we take the representation matrices as $(T^I)^{}_{JK} = -{\rm i} \epsilon^{IJK}$ (for $I, J, K = 1, 2, 3$),
where $\epsilon^{IJK}$ is the totally antisymmetric Levi-Civita tensor.
As for the adjoint representation of the ${\rm SU}(3)^{}_{\rm c}$ group, $T^A = \lambda^A/2$ is defined with $\lambda^A$ being the Gell-Mann matrices (for $A = 1, 2, \cdots, 8$). In the subsequent discussions, we rewrite $\Sigma_{\rm R}^{}$ in the adjoint representation as $\Sigma_{\rm R}^{} \equiv \sigma^I \cdot \mathrm{\Sigma}_{\rm R}^I/\sqrt2$ and introduce the three-vector $\mathbf{\Sigma}^{}_{\rm R} \equiv (\Sigma^1_{\rm R}, \Sigma^2_{\rm R}, \Sigma^3_{\rm R})$ in the weak-isospin space. Hence the Lagrangian in Eq.~\eqref{eq:Luv} can be recast into
\bea
\mc{L}^{}_{\rm type-III} &=& \mc{L}^{}_{\rm SM} + {\overline{\mathbf{\Sigma}}^{}_{\rm R}} \cdot \rmI \slashed{D} \mathbf{\Sigma}^{}_{\rm R} - \left({\hat{\mathbf{Y}}}^{}_\mathrm{\Sigma} \cdot\mathbf{\Sigma}^{}_{\rm R} + \frac{{ {M}}^{}_{\Sigma}}{2} {\overline{\mathbf{\Sigma}_{\rm R}^{\rm c}}} \cdot \mathbf{\Sigma}^{}_{\rm R} + \text{h.c.}\right) \nonumber \\
&=& \mc{L}^{}_{\rm SM} + \frac{1}{2}\overline{\mathbf{\Sigma}} \cdot \rmI \slashed{D}\mathbf{\Sigma} - \frac{{ {M}}^{}_{\Sigma}}{2} \overline{\mathbf{\Sigma}} \cdot \mathbf{\Sigma} -    \bigg( \frac{\hat{\mathbf{Y}}^{}_{\Sigma} + {\hat{\mathbf{Y}}_{\Sigma}^{\rm c \dag}} }{2} \cdot \mathbf{\Sigma} + \overline{\mathbf{\Sigma}} \cdot \frac{{\hat{\mathbf{Y}}}_{\Sigma}^{\rm c} + {\hat{\mathbf{Y}}}_{\Sigma}^\dag}{2} \bigg) \;,
\label{eq:Luv2}
\eea
in which $\mathbf{\Sigma} = \mathbf{\Sigma}^{}_{\rm R} + \mathbf{\Sigma}_{\rm R}^{\rm c}$ has been defined, and $\text{Tr}\left(\overline{ {\Sigma}} {\Sigma}^{\rm c}\right) = \text{Tr}\left(\overline{ {\Sigma}} {\Sigma}\right) = \overline{\mathbf{\Sigma}}\cdot\mathbf{\Sigma}$ has been used. It is worth stressing that all quantities in boldface represent the vectors in the weak-isospin space, and their components are defined as $\mathbf{\Sigma}^I = \Sigma^I$, $({\hat{\mathbf{Y}}}_\mathrm{\Sigma})^I = \overline{\ell^{}_{\rm L}} Y^{}_{\Sigma} \sigma^I \widetilde{H}$ and $({\hat{\mathbf{Y}}}_{\Sigma}^{\rm c})^I = H^\dag Y_\mathrm{\Sigma}^{\rm T} \sigma^I\widetilde{\ell_{\rm L}^{}}$, with $I = 1, 2, 3$ and $\widetilde{\ell_{\rm L}^{}}={\rm i}\sigma^2 \ell_{\rm L}^{\rm c}$. In these notations, the dot in Eq.~\eqref{eq:Luv2} is just the inner product between two three-vectors. With the Lagrangian in Eq.~(\ref{eq:Luv2}), it is straightforward to derive the equation of motion (EOM) for $\mathbf{\Sigma}$, i.e.,
\bea
\rmI \slashed{D}\mathbf{\Sigma} - { {M}}^{}_\mathrm{\Sigma} \mathbf{\Sigma} - \left({\hat{\mathbf{Y}}}_\mathrm{\Sigma}^\dag +{\hat{\mathbf{Y}}}_\mathrm{\Sigma}^{\rm c}\right) = 0 \;.
\label{eq:SigmaEOM}
\eea
By solving the above EOM, one can find the classical field $\mathbf{\Sigma}^{}_{\rm c}$ and then expand it with respect to $1/M_\Sigma^{}$ in order to obtain the local field $\hat{\mathbf{\Sigma}}^{}_{\rm c}$. Integrating out the heavy field $\mathbf{\Sigma}$ at the tree level is equivalent to substituting the local classical field $\hat{\mathbf{\Sigma}}^{}_{\rm c}$ back into the Lagrangian of the UV theory. Since we are interested in the operators in the SEFT-III up to dim-6, it is sufficient to retain the terms up to $\mc{O}(1/{ {M}}_{\Sigma}^2)$ in the local field, i.e.,
\bea
\hat{\mathbf{\Sigma}}^{}_{\rm c} &=& \frac{1}{\rmI \slashed{D} - { {M}}^{}_{\Sigma}} \left({\hat{\mathbf{Y}}}_{\Sigma}^\dag + {\hat{\mathbf{Y}}}_{\Sigma}^{\rm c}\right) = -\frac{1}{{ {M}}^{}_{\Sigma}} \left(1 + \frac{\rmI \slashed{D}}{{ {M}}^{}_{\Sigma}}\right) \left({\hat{\mathbf{Y}}}_{\Sigma}^\dag + {\hat{\mathbf{Y}}}_{\Sigma}^{\rm c}\right) + \mc{O}\left(\frac{1}{{ {M}}_{\Sigma}^3}\right)\;.
\label{eq:classicalf}
\eea

After inserting the local classical field $\hat{\mathbf{\Sigma}}^{}_{\rm c}$ in Eq.~\eqref{eq:classicalf} into the full Lagrangian Eq.~\eqref{eq:Luv2}, we complete the tree-level matching of the type-III seesaw model onto the SMEFT with the dim-5 and dim-6 operators. Some comments on the results of this tree-level matching are in order.
\begin{itemize}
\item The tree-level Lagrangian of the SEFT-III turns out to be $
\mc{L}_\text{SEFT-III}^{\rm tree} = \mc{L}^{}_{\rm SM} -(1/2) ({\hat{\mathbf{Y}}}_{\Sigma}+{{\hat{\mathbf{Y}}}_{\Sigma}^{\rm c \dag}})\cdot \hat{\mathbf{\Sigma}}^{}_{\rm c}$, where $\hat{\mathbf{\Sigma}}^{}_{\rm c}$ is given in Eq.~\eqref{eq:classicalf}. Keeping the terms of $\mc{O}(1/M^{}_{\Sigma})$ in Eq.~\eqref{eq:classicalf}, we get the dim-5 Weinberg operator
\bea
\mc{L}^{\rm tree}_\text{dim-5} = \frac{{\hat{\mathbf{Y}}}^{}_\mathrm{\Sigma}}{2} \frac{1}{{ {M}}^{}_\mathrm{\Sigma}}{\hat{\mathbf{Y}}}_\mathrm{\Sigma}^{\rm c} + {\rm h.c.} = \frac{1}{2}\left(Y^{}_\mathrm{\Sigma} \frac{1}{{ {M}}^{}_\mathrm{\Sigma}} Y_\mathrm{\Sigma}^{\rm T}\right)^{\alpha \beta} \overline{\ell^{\alpha}_{\rm L}} \widetilde{H}{\widetilde{H}}^{\rm T} \ell_{\rm L}^{\beta \rm{c}} + \text{h.c.} \equiv \big( C^{(5)}_{\rm eff} \big)^{\alpha\beta}_{\rm tree} \mc{O}^{\alpha \beta}_5 + {\rm h.c.} \;, \quad
\label{eq:dim5-tree}
\eea
where the identity $\sigma_{ab}^I \sigma_{cd}^I = 2\delta^{}_{ad} \delta^{}_{bc} - \delta^{}_{ab} \delta^{}_{cd}$ has been used and $H^\dag\widetilde{H}=H^\dag\epsilon H^\ast=0$ is implied. One can see that the Weinberg operator $\mc{O}^{\alpha \beta}_5 \equiv \overline{\ell^{\alpha}_{\rm L}} \widetilde{H}{\widetilde{H}}^{\rm T} \ell_{\rm L}^{\beta \rm{c}}$ is successfully derived with the Wilson coefficient $\big( C^{(5)}_{\rm eff} \big)_{\rm tree} = (1/2) Y^{}_\mathrm{\Sigma} M^{-1}_\mathrm{\Sigma} Y_\mathrm{\Sigma}^{\rm T}$.

\item In a similar way, to obtain the dim-6 operators, one should take out the terms of order $1/M_\Sigma^2$ in Eq.~\eqref{eq:classicalf}. More explicitly, the dim-6 operator in the SEFT-III at the tree level reads
\bea
\mc{L}^{\rm tree}_\text{dim-6}&=& \frac{{\hat{\mathbf{Y}}}^{}_\mathrm{\Sigma}}{2} \frac{\rmI \slashed{D}}{{ {M}}_{\Sigma}^2} {\hat{\mathbf{Y}}}_{\Sigma}^\dag
+ \frac{{{\hat{\mathbf{Y}}}_{\Sigma}^{\rm c\dag}}}{2} \frac{\rmI \slashed{D}}{{ {M}}_\mathrm{\Sigma}^2} {\hat{\mathbf{Y}}}_\mathrm{\Sigma}^{\rm c}
= \left(Y^{}_\mathrm{\Sigma}\frac{1}{{ {M}}_\mathrm{\Sigma}^2} Y_\mathrm{\Sigma}^\dag\right) \cdot \left(\overline{\ell^{}_{\rm L}}\sigma^I \widetilde{H}\right) \rmI \slashed{D} \left({\widetilde{H}}^\dag \sigma^I \ell^{}_{\rm L}\right)\;,
\eea
which coincides the result in Ref.~\cite{Abada:2007ux}. Converting the above dim-6 operator into those in the Warsaw basis, we get the following three operators
\bea
O^{(1)}_{H\ell} &=& (H^\dag \rmI \overleftrightarrow{D}^{}_\mu H) \left(\overline{\ell^{}_{\rm L}} \gamma^\mu \ell^{}_{\rm L} \right) \;, O^{(3)}_{H\ell} = (H^\dag \rmI \overleftrightarrow{D}_\mu^I H) \left(\overline{\ell^{}_{\rm L}} \sigma^I \gamma^\mu \ell^{}_{\rm L}\right) \;, O^{}_{eH} = ( H^\dagger H) \left( \overline{\ell^{}_{\rm L}} E^{}_{\rm R} H \right)\;, \quad
\eea
with their tree-level Wilson coefficients
\bea
C^{(1)}_{H\ell\text{-tree}}&=& \frac{3}{4} Y_\Sigma^{}M_\Sigma^{-2} Y_\Sigma^\dag \;,\quad C^{(3)}_{H\ell \text{-tree}} = \frac{1}{4} Y_\Sigma^{}M_\Sigma^{-2} Y_\Sigma^\dag \;, \quad C^{}_{eH \text{-tree}} = Y_\Sigma^{}M_\Sigma^{-2} Y_\Sigma^\dag  Y_l^{} \;.
\label{eq:treematching}
\eea
Note that the lepton-flavor indices of these dim-6 operators and their Wilson coefficients have been suppressed.
\end{itemize}

Thus far we have matched the type-III seesaw model onto the SMEFT at the tree level. The unique dim-5 Weinberg operator and three dim-6 operators in the Warsaw basis, together with their Wilson coefficients in terms of the model parameters in the full theory, are derived. The technical details of the one-loop matching have been clearly explained and extensively applied to various UV theories in Refs.~\cite{Beneke:1997zp,Jantzen:2011nz,Zhang:2016pja,Jiang:2018pbd, Dittmaier:2021fls, Fuentes-Martin:2020udw,Carmona:2021xtq,Fuentes-Martin:2022jrf}. Therefore, we shall just summarize the loop-level matching results in next section.

\section{One-loop Matching}
\label{result}
The publicly available packages \texttt{Matchmakereft}~\cite{Carmona:2021xtq} and \texttt{Matchete}~\cite{Fuentes-Martin:2022jrf} have been designed to accomplish the one-loop matching of any renormalizable UV model onto the SMEFT, based on the diagrammatic approach and the functional approach, respectively. We have utilized these two packages to carry out the one-loop matching of the type-III seesaw model and made a careful cross-check to ensure the correctness of the matching results. In the following, we will list the complete one-loop matching results, including threshold corrections to the SM couplings, the matching conditions for the Weinberg operator and for the dim-6 operators in the Warsaw basis.

\subsection{Threshold corrections}
The renormalizable terms of dim-4 that already exist in the SM receive threshold corrections from the one-loop matching. We collect these corrections to the SM Lagrangian as below
\begin{eqnarray}\label{eq:correction-GB}
\delta \mathcal{L}^{}_{\rm SM} &=& \delta Z^{}_{G} G^A_{\mu\nu} G^{A\mu\nu} + \delta Z^{}_W W^I_{\mu\nu} W^{I\mu\nu} + \delta Z^{}_B B^{}_{\mu\nu} B^{\mu\nu} + \sum^{}_f  \overline{f} \delta Z^{}_f \rmI \slashed{D} f
	\nonumber
	\\
&&+ \left( \overline{Q^{}_{\rm}} \delta Y^{}_{\rm u} \widetilde{H} U^{}_{\rm R} + \overline{Q^{}_{\rm L}} \delta Y^{}_{\rm d} H D^{}_{\rm R} + \overline{\ell^{}_{\rm L}} \delta Y^{}_l H E^{}_{\rm R} + {\rm h.c.} \right)
	\nonumber
	\\
	&& + \delta Z^{}_H \left( D^{}_\mu H \right)^\dagger \left( D^\mu H \right) + \delta m^2_H H^\dagger H + \delta \lambda \left( H^\dagger H \right)^2 \;.
	\label{eq:threshold}
\end{eqnarray}
As the kinetic terms for the Higgs doublet, gauge bosons and fermions are modified, one has to normalize these fields to maintain the canonical form. For the Higgs doublet and fermions, we redefine $H \to (1 - \delta Z^{}_H/2)H$ and $f\to (1-\delta Z_f^{}/2)f$. For the gauge bosons, one has to redefine the gauge-boson fields and gauge couplings $g^{}_1$ and $g^{}_2$ simultaneously, i.e.,
\begin{eqnarray}
    \begin{cases}
    B^{}_\mu \to \left( 1 + 2\delta Z^{}_B \right) B^{}_\mu
    \\
    g^{}_1 \to g^{\rm eff}_1 = \left( 1 - 2\delta Z^{}_B \right) g^{}_1
    \end{cases}
    \;,\qquad
    \begin{cases}
    W^I_\mu \to \left( 1 + 2\delta Z^{}_W \right) W^I_\mu
    \\
    g^{}_2 \to g^{\rm eff}_2 = \left( 1 - 2\delta Z^{}_W \right) g^{}_2
    \end{cases}
    \;,
    \label{eq:gauge-boson}
\end{eqnarray}
to keep the covariant derivative $D^{}_\mu$ intact. For notational simplicity, we introduce the effective parameter $g^{}_{\rm eff}$ and the threshold correction $\delta g^{}_{\rm eff}$ such that $g^{}_{\rm eff} = g + \delta g^{}_{\rm eff}$, where $g$ is the original parameter belonging to $\{m^2, \lambda, Y^{}_l, Y^{}_{\rm u}, Y^{}_{\rm d}, g^{}_2\}$. All the threshold corrections are
\bea
\delta g^{}_{2,\rm eff} &=& -2 \delta Z^{}_W g^{}_2 = \frac{g_2^3}{24\pi^2}\tr\left(L^{}_\Sigma\right) \;, \label{eq:eff-g-1} \\
\delta m^2_{\rm eff} &=& - m^2 \delta Z^{}_H - \delta m^2 \nonumber \\
&=& \frac{m^4}{16\pi^2}\tr\left(Y_\Sigma^{} M_\Sigma^{-2} Y_\Sigma^\dag \right) + \frac{3m^2}{32\pi^2} \tr\left(Y^{}_\Sigma \left(1+2L^{}_\Sigma\right) Y_\Sigma^\dag\right) -\frac{3}{8\pi^2} \tr\left(Y_\Sigma^{} M_\Sigma^2 \left(1+L^{}_\Sigma\right) Y_\Sigma^\dag\right)  \;, \\
\delta Y_{\rm u, eff}^{} &=& -Y^{}_{\rm u} \delta Z^{}_H/2 - \delta Y^{}_{\rm u} = \frac{m^2 Y_{\rm u}^{\alpha\beta}}{16\pi^2}\tr\left(Y_\Sigma^{} M_\Sigma^{-2} Y_\Sigma^\dag\right) + \frac{3Y_{\rm u}^{\alpha\beta}}{64\pi^2} \tr\left(Y^{}_\Sigma \left(1+2L^{}_\Sigma\right) Y_\Sigma^\dag\right) \;, \\
\delta Y_{\rm d, eff}^{} &=& - Y^{}_{\rm d} \delta Z^{}_H/2 - \delta Y^{}_{\rm d} = \frac{m^2 Y_{\rm d}^{\alpha\beta}}{16\pi^2} \tr\left(Y_\Sigma^{} M_\Sigma^{-2} Y_\Sigma^\dag\right) + \frac{3Y_{\rm d}^{\alpha\beta}}{64\pi^2} \tr\left(Y^{}_\Sigma \left(1+2L^{}_\Sigma\right) Y_\Sigma^\dag\right)   \;, \\
\delta Y_{l,{\rm eff}}^{} &=& -Y^{}_{l} \delta Z^{}_H/2 - \frac{1}{2}\delta Z^{}_{\ell}Y^{}_{l} - \delta Y^{}_{l} = \frac{m^2 Y_l^{\alpha\beta}}{16\pi^2} \tr\left(Y_\Sigma^{} M_\Sigma^{-2} Y_\Sigma^\dag\right) + \frac{9m^2}{64\pi^2} \left(Y_\Sigma^{} M_\Sigma^{-2} \left(3+2L^{}_\Sigma\right) Y_\Sigma^\dag Y_l^{}\right)^{\alpha\beta} \nonumber \\
&& +\frac{3Y_l^{\alpha\beta}}{64\pi^2} \tr\left(Y^{}_\Sigma \left(1+2L^{}_\Sigma\right) Y_\Sigma^\dag\right) + \frac{3}{128\pi^2} \left(Y^{}_\Sigma\left(11+10L^{}_\Sigma\right) Y_\Sigma^\dag Y_l^{}\right)^{\alpha\beta}  \;, \\
\delta\lambda^{}_{\rm eff} &=& - 2\lambda \delta Z^{}_H - \delta \lambda \nonumber \\
&=& \frac{m^2 g_2^2}{96\pi^2} \tr \bigg[
	Y^{}_\Sigma \frac{(7+2L^{}_\Sigma)}{M_\Sigma^2}  Y_\Sigma^\dag
\bigg] - \frac{m^2 g_2^4}{120\pi^2} \tr\left(M_\Sigma^{-2}\right)
\nonumber \\
&& +\frac{\lambda}{16\pi^2} \left\{4m^2\tr\left(Y_\Sigma^{}M_\Sigma^{-2}Y_\Sigma^\dag\right) + \tr\left[Y^{}_\Sigma\left(3+6L^{}_\Sigma\right)Y_\Sigma^\dag\right] \right\} \nonumber \\
&& -\frac{m^2}{32\pi^2} \tr\left[Y_\Sigma^{} M_\Sigma^{-2} \left(7-2L^{}_\Sigma\right) Y_\Sigma^\dag Y_l^{} Y_l^{\dag}\right] -\frac{1}{4\pi^2} \tr\left[Y^{}_\Sigma \left(1+L^{}_\Sigma\right) Y_\Sigma^\dag Y_l^{}Y_l^\dag\right] \nonumber \\
&& -\frac{5\left(Y_\Sigma^\dag Y_\Sigma^{}\right)_{ik}\left(Y_\Sigma^\dag Y_\Sigma^{}\right)_{ki}}{16\pi^2\left(M_i^2-M_k^2\right)} \bigg[M_i^2\left(1+L^{}_i\right)-M_k^2\left(1+L^{}_k\right) \bigg] + \frac{\left(Y_\Sigma^\dag Y_\Sigma^{}\right)^{}_{ik} \left(Y_\Sigma^\dag Y_\Sigma^{}\right)^{}_{ik} M^{}_i M^{}_k L^{}_{ik}}{16\pi^2\left(M_i^2-M_k^2\right)} \nonumber \\
&& - \frac{m^2\left(Y_\Sigma^\dag Y_\Sigma^{}\right)^{}_{ik} \left(Y_\Sigma^\dag Y_\Sigma^{}\right)^{}_{ki}}{16\pi^2\left(M_i^2-M_k^2\right)^3} \bigg[M_i^4\left(4+3L^{}_{ik}\right) - M_k^4 \left(4-3L^{}_{ik}\right) - 14M_i^2 M_k^2 L^{}_{ik}\bigg] \nonumber \\
&& - \frac{m^2\left(Y_\Sigma^\dag Y_\Sigma^{}\right)^{}_{ik} \left(Y_\Sigma^\dag Y_\Sigma^{}\right)^{}_{ik}} {32\pi^2 M^{}_i M^{}_k \left(M_i^2 - M_k^2\right)^3} \bigg[ M_i^6 \left(1+2L^{}_k\right) -M_k^6 \left(1+2L^{}_i\right)   \nonumber\\
&&   - M_i^4 M_k^2 \left(19+10L^{}_i -4L^{}_k\right) +M_k^4 M_i^2 \left(19+10L^{}_k-4L^{}_i\right) \bigg]  \;, \label{eq:eff-g-n}
\eea
where ``tr'' indicates the trace operation in the flavor space, and
\bea
L^{}_\Sigma \equiv \text{diag}\bigg\{\log{\left(\frac{\mu^2}{M_1^2}\right)}, \log{\left(\frac{\mu^2}{M_2^2}\right)}, \log{\left(\frac{\mu^2}{M_3^2}\right)}\bigg\} \equiv \text{diag}\left\{L^{}_1, L^{}_2, L^{}_3 \right\}\;,
\eea
with $\mu$ being the 't Hooft mass scale, and $L^{}_{ik} \equiv \log{\left( M_i^2/M_k^2\right)}$ has been defined. Note that the fermion triplet has $Y = 0$, so no threshold correction to the gauge coupling $g^{}_1$ appears at one-loop.

\subsection{Matching condition for the dim-5 operator}
The Wilson coefficient of the dim-5 operator at the one-loop level is given by
\bea
\big( C^{(5)}_{\rm eff} \big)^{\alpha\beta}_{\rm loop} &=& +\frac{g_1^2}{128\pi^2}\left[Y_\Sigma^{}\frac{\left(1+3L_\Sigma\right)}{M_\Sigma}Y_\Sigma^{\rm T}\right]^{\alpha\beta}+\frac{g_2^2}{128\pi^2}\left[Y_\Sigma^{}\frac{\left(1+3L_\Sigma\right)}{M_\Sigma}Y_\Sigma^{\rm T}\right]^{\alpha\beta}\nonumber \\
&& +\frac{\lambda}{16\pi^2}\left[Y_\Sigma^{}\frac{\left(1+L_\Sigma\right)}{M_\Sigma}Y_\Sigma^{\rm T}\right]^{\alpha\beta}-\frac{1}{16\pi^2}\left[Y_l^{} Y_l^\dag Y_\Sigma^{} \frac{\left(1+L_\Sigma\right)}{M_\Sigma}Y_\Sigma^{\rm T} +{\rm Trans.} \right]^{\alpha\beta} \nonumber \\
&& -\frac{3}{256\pi^2} \bigg[Y_\Sigma^{}\left(3+2L_\Sigma \right)Y_\Sigma^{\rm T} Y_\Sigma^{\ast}  M^{-1}_\Sigma Y_\Sigma^{\rm T} +{\rm Trans.}  \bigg]^{\alpha \beta } \nonumber \\
&& -\frac{3}{64\pi^2} \tr \left[ Y_\Sigma^{}\left(1+2L_\Sigma^{}\right) Y_\Sigma^\dag\right]\left(Y_\Sigma^{} M^{-1}_\Sigma Y_\Sigma^{\rm T}\right)^{\alpha \beta}  \;,
\label{eq:dim5-loop}
\eea
where ``Trans.'' denotes the transpose of the preceding term.

\subsection{Matching conditions for the dim-6 operators}
For different types of dim-6 operators,  the matching conditions for their Wilson coefficients at the one-loop level are summarized as follows
\begin{table}
\begin{center}
	\hspace{0.05cm}
	   \renewcommand\arraystretch{1.7}
		  \scalebox{0.66}
		  {
			\begin{tabular}{||c|c||c|c||c|c||}
				\hline \hline \multicolumn{2}{||c||}{$X^{3}$} & \multicolumn{2}{c||}{$H^{6} \text { and }
			H^{4} D^{2}$} & \multicolumn{2}{c||}{$\psi^{2} H^{3}$} \\
			\hline $O_{G}$ & $f^{A B C} G_{\mu}^{A \nu} G_{\nu}^{B \rho} G_{\rho}^{C \mu} $
			& \cellcolor{gray!30}{$\bm{O_{H}}$} & \cellcolor{gray!30}{$\bm{\left(H^{\dagger} H\right)^{3}}$}
			& \cellcolor{gray!30}{$\bm{O_{e H}^{\alpha \beta }}$}
			& \cellcolor{gray!30}{$\bm{\left(H^{\dagger} H\right)\left(\overline{{\ell}^{}_{\alpha \text{L}}} E_{\beta\text{R}} H\right)}$} \\
			$O_{\widetilde{G}}$ & $f^{A B C} \widetilde{G}_{\mu}^{A \nu} G_{\nu}^{B \rho}
			G_{\rho}^{C \mu}$ & \cellcolor{gray!30}{$\bm{O_{H \square}}$} & \cellcolor{gray!30}{$\bm{\left(H^{\dagger} H\right)
			\square\left(H^{\dagger} H\right)}$} & \cellcolor{gray!30}{$\bm{O_{u H}^{\alpha \beta }}$} & \cellcolor{gray!30}
			{$\bm{\left(H^{\dagger} H\right)\left(\overline{Q_{\alpha \text{L}}} {U}^{}_{\beta\text{R}} \widetilde{H}\right)}$} \\
			\cellcolor{gray!65}{$\bm{O_{W}}$} & \cellcolor{gray!65}{$\bm{\epsilon^{I J K} W_{\mu}^{I \nu} W_{\nu}^{J \rho} W_{\rho}^{K \mu}}$}
			 & \cellcolor{gray!30}{$\bm{O_{H D}}$}
			 & \cellcolor{gray!30}{$\bm{\left(H^{\dagger} D^{\mu} H\right)^{\star}\left(H^{\dagger} D_{\mu}
			 H\right)}$} & \cellcolor{gray!30}{$\bm{O_{d H}^{\alpha \beta }} $}& \cellcolor{gray!30}{$\bm{\left(H^{\dagger} H\right)
			 \left(\overline{Q_{\alpha \text{L}}} D_{\beta\text{R}} H\right)}$}\\
			$O_{\widetilde{W}}$ & $\epsilon^{I J K} \widetilde{W}_{\mu}^{I \nu}
			W_{\nu}^{J \rho} W_{\rho}^{K \mu}$ & & & & \\
			\hline \hline \multicolumn{2}{||c||}{$X^{2} H^{2}$} & \multicolumn{2}{c||}{$\psi^{2} X H$} & \multicolumn{2}
			{|c||}{$\psi^{2} H^{2} D$} \\
			\hline $O_{H G} $&$ H^{\dagger} H G_{\mu \nu}^{A} G^{A \mu \nu}$
			 & \cellcolor{gray!30}{$\bm{O_{e W}^{\alpha \beta }}$} &\cellcolor{gray!30}{ $\bm{\left(\overline{{\ell}^{}_{\alpha \text{L}}} \sigma^{\mu \nu} E_{\beta\text{R}}\right) \sigma^{I} H
			 W_{\mu \nu}^{I}}$} & \cellcolor{gray!30}{$\bm{O_{H l}^{(1)\alpha\beta}}$} & \cellcolor{gray!30}{$\bm{(H^{\dagger} \text{i}
			 \overleftrightarrow{D}_{\mu} H)\left(\overline{{\ell}^{}_{\alpha \text{L}}} \gamma^{\mu}
			 {\ell}^{}_{\beta\text{L}}\right)}$} \\
		   $ O_{H \widetilde{G}}$ & $H^{\dagger} H \widetilde{G}_{\mu \nu}^{A}
			G^{A \mu \nu}$ & \cellcolor{gray!30}{$\bm{O_{e B}^{\alpha \beta }}$} & \cellcolor{gray!30}{$\bm{\left(\overline{{\ell}^{}_{\alpha \text{L}}} \sigma^{\mu \nu} E_{\beta\text{R}}\right) H
			B_{\mu \nu}}$} & \cellcolor{gray!30}{$\bm{O_{H l}^{(3)\alpha \beta }}$} & \cellcolor{gray!30}
			{$\bm{(H^{\dagger} \text{i} \overleftrightarrow{D}_{\mu}^IH)\left(\overline{{\ell}^{}_{\alpha \text{L}}} \sigma^{I} \gamma^{\mu} {\ell}^{}_{\beta\text{L}}\right)}$} \\
			\cellcolor{gray!30}{$\bm{O_{H W}}$}& \cellcolor{gray!30}{$\bm{H^{\dagger} H W_{\mu \nu}^{I} W^{I \mu \nu}}$}
			& $O_{u G}^{\alpha \beta }$ & $ \left(\overline{Q_{\alpha \text{L}}} \sigma^{\mu \nu} T^{A} {U}^{}_{\beta\text{R}}\right) \widetilde{H} G_{\mu \nu}^{A} $
			& \cellcolor{gray!30}{$\bm{O_{H e}^{\alpha \beta }}$} & \cellcolor{gray!30}{$\bm{(H^{\dagger} \text{i} \overleftrightarrow{D}_{\mu}
			H)\left(\overline{E_{\alpha\text{R}}} \gamma^{\mu} E_{\beta\text{R}}\right)}$} \\
		   $ O_{H \widetilde{W}}$ & $H^{\dagger} H \widetilde{W}_{\mu \nu}^{I}
			W^{I \mu \nu}$ & $O_{u W}^{\alpha \beta }$ & $\left(\overline{Q_{\alpha \text{L}} }\sigma^{\mu \nu} {U}^{}_{\beta\text{R}}\right) \sigma^{I}
			\widetilde{H} W_{\mu \nu}^{I}$ & \cellcolor{gray!30}{$\bm{O_{H q}^{(1)\alpha \beta }}$} & \cellcolor{gray!30}{$\bm{(H^{\dagger}
			\text{i} \overleftrightarrow{D}_{\mu} H)\left(\overline{Q_{\alpha \text{L}}} \gamma^{\mu} {Q}^{}_{\beta\text{L}}\right)}$} \\
			\cellcolor{gray!30}{$\bm{O_{H B}}$} & \cellcolor{gray!30}{$\bm{H^{\dagger} H B_{\mu \nu} B^{\mu \nu}}$} & $O_{u B}^{\alpha \beta } $
			&$ \left(\overline{Q_{\alpha \text{L}}} \sigma^{\mu \nu} {U}^{}_{\beta \text{R}}\right) \widetilde{H} B_{\mu \nu} $
			& \cellcolor{gray!30}{$\bm{O_{H q}^{(3)\alpha \beta }}$ }& \cellcolor{gray!30}{$\bm{(H^{\dagger} \text{i} \overleftrightarrow{D}_{\mu}^{I}
			H)\left(\overline{Q_{\alpha \text{L}}} \sigma^{I} \gamma^{\mu} {Q}^{}_{\beta\text{L}}\right) }$}\\
			$O_{H \widetilde{B}}$ & $H^{\dagger} H \widetilde{B}_{\mu \nu}
			B^{\mu \nu}$ & $O_{d G}^{\alpha \beta }$& $\left(\overline{Q_{\alpha \text{L}}} \sigma^{\mu \nu} T^{A} D_{\beta\text{R}}\right)
			H G_{\mu \nu}^{A}$ & \cellcolor{gray!30}{$\bm{O_{H u}^{\alpha \beta }}$} & \cellcolor{gray!30}
			{$\bm{(H^{\dagger} \text{i} \overleftrightarrow{D}_{\mu} H)\left(\overline{{U}^{}_{\alpha \text{R}}} \gamma^{\mu} {U}^{}_{\beta \text{R}}\right)}$} \\
			\cellcolor{gray!30}{$\bm{O_{H W B}}$} & \cellcolor{gray!30}{$\bm{H^{\dagger} \sigma^{I} H W_{\mu \nu}^{I} B^{\mu \nu}}$}
			 & $O_{d W}^{\alpha \beta } $
			& $\left(\overline{Q_{\alpha \text{L}}} \sigma^{\mu \nu} D_{\beta\text{R}}\right) \sigma^{I} H W_{\mu \nu}^{I}$ & \cellcolor{gray!30}{$\bm{O_{H d}^{\alpha \beta }}$}
			& \cellcolor{gray!30}{$\bm{(H^{\dagger} \text{i} \overleftrightarrow{D}_{\mu} H)\left(\overline{D_{\alpha \text{R}}}
			\gamma^{\mu} D_{\beta\text{R}}\right)}$} \\
		   $ O_{H \widetilde{W} B} $& $H^{\dagger} \sigma^{I} H \widetilde{W}_{\mu \nu}^{I}
			B^{\mu \nu} $& $O_{d B}^{\alpha \beta } $& $\left(\overline{Q_{\alpha \text{L}}} \sigma^{\mu \nu} D_{\beta\text{R}}\right) H B_{\mu \nu}$
			 & $O_{H u d}^{\alpha \beta }$ & $\text{i}(\widetilde{H}^{\dagger} D_{\mu}
			 H)\left(\overline{{U}^{}_{\alpha \text{R}}} \gamma^{\mu} D_{\beta\text{R}}\right)$ \\
			\hline \hline \multicolumn{2}{||c||}
			{$\overline{{\text{L}}}\text{L}\overline{{\text{L}}^{}}\text{L}$} & \multicolumn{2}{c||} {$\overline{\text{R}}\text{R}\overline{\text{R}}\text{R}$} & \multicolumn{2}{c||}{$\overline{{\text{L}}^{}}\text{L}\overline{\text{R}}\text{R}$} \\
			 \hline \rowcolor{gray!30} $\bm{O_{\ell\ell}^{\alpha \beta \gamma \delta} }$& $\bm{\left({\overline{{\ell}^{}_{\alpha \text{L}}}}
			 \gamma^\mu {\ell}^{}_{\beta\text{L}}\right)\left({\overline{{\ell}^{}_{\gamma \text{L}}}}\gamma^\mu {\ell}^{}_{\delta\text{L}}\right)}$
			&\cellcolor{gray!65} $O_{ee}^{\alpha \beta \gamma \delta}$ &\cellcolor{gray!65} ${\left({\overline{E_{\alpha\text{R}}}}\gamma^\mu E_{\beta\text{R}}\right)
			\left({\overline{E_{\gamma\text{R}}}}\gamma^\mu E_{\delta\text{R}}\right) }$&$\bm{ O_{\ell e}^{\alpha \beta \gamma \delta}}$
			&$\bm{ \left({\overline{{\ell}^{}_{\alpha \text{L}}}}\gamma^\mu {\ell}^{}_{\beta\text{L}}\right)
			\left({\overline{E_{\gamma\text{R}}}}\gamma^\mu E_{\delta\text{R}}\right)}$ \\
			\cellcolor{gray!65} {$O_{qq}^{(1) \alpha \beta \gamma \delta} $}&\cellcolor{gray!65} $ {\left({\overline{Q_{\alpha \text{L}}}}
			\gamma^\mu {Q}^{}_{\beta \text{L}}\right)\left({\overline{Q_{\gamma \text{L}}}}\gamma^\mu {Q}^{}_{\delta\text{L}}\right)}$
        &\cellcolor{gray!65} $O_{uu}^{\alpha \beta \gamma \delta} $&\cellcolor{gray!65} $ {\left({\overline{{U}^{}_{\alpha \text{R}}}}\gamma^\mu
        {U}^{}_{\beta\text{R}}\right)\left({\overline{{U}^{}_{\gamma \text{R}}}}\gamma^\mu {U}^{}_{\delta \text{R}}\right)}$ &\cellcolor{gray!30} $ \bm{O_{\ell u}^{\alpha \beta \gamma \delta} }$
        &\cellcolor{gray!30} $\bm{\left({\overline{{\ell}^{}_{\alpha \text{L}}}}\gamma^\mu {\ell}^{}_{\beta\text{L}}\right)\left({\overline{{U}^{}_{\gamma \text{R}}}}\gamma^\mu {U}^{}_{\delta\text{R}}\right)}$\\
        \rowcolor{gray!30} \cellcolor{gray!65} $\bm{O_{qq}^{(3)\alpha \beta \gamma \delta}}$ & \cellcolor{gray!65} $\bm{ {\left({\overline{Q_{\alpha \text{L}}}}
        \sigma^I\gamma^\mu {Q}^{}_{\beta\text{L}}\right)\left({\overline{Q_{\gamma \text{L}}}}
        \sigma^I\gamma^\mu {Q}^{}_{\delta\text{L}}\right) }}$&\cellcolor{gray!65}$ O_{dd}^{\alpha \beta \gamma \delta}$
        &\cellcolor{gray!65} $ {\left({\overline{D_{\alpha\text{R}}}}\gamma^\mu D_{\beta\text{R}}\right)
        \left({\overline{D_{\gamma\text{R}}}}\gamma^\mu D_{\delta\text{R}}\right)}$ & $\bm{O_{\ell d}^{\alpha \beta \gamma \delta}}$
        & $\bm{\left({\overline{{\ell}^{}_{\alpha \text{L}}}}\gamma^\mu {\ell}^{}_{\beta\text{L}}\right)
        \left({\overline{D_{\gamma\text{R}}}}\gamma^\mu D_{\delta\text{R}}\right) }$\\
        \rowcolor{gray!30} $\bm{O_{\ell q}^{(1)\alpha \beta \gamma \delta}} $& $\bm{\left({\overline{{\ell}^{}_{\alpha \text{L}}}}
        \gamma^\mu {\ell}^{}_{\beta\text{L}}\right)\left({\overline{Q_{\gamma \text{L}}}}\gamma^\mu {Q}^{}_{\delta\text{L}}\right)}$ & \cellcolor{gray!65}  $O_{eu}^{\alpha \beta \gamma \delta}$
        &  \cellcolor{gray!65} $ {\left({\overline{E_{\alpha\text{R}}}}\gamma^\mu E_{\beta\text{R}}\right)\left({\overline{{U}^{}_{\gamma \text{R}}}}\gamma^\mu {U}^{}_{\delta\text{R}}\right) }$& \cellcolor{gray!65}  $O_{qe}^{\alpha \beta \gamma \delta}$
        & \cellcolor{gray!65} $ {\left({\overline{Q_{\alpha \text{L}}}}\gamma^\mu {Q}^{}_{\beta\text{L}}\right)\left({\overline{E_{\gamma\text{R}}}}\gamma^\mu E_{\delta\text{R}}\right)}$ \\
        \rowcolor{gray!30} $\bm{O_{\ell q}^{(3)\alpha \beta \gamma \delta}}$ &$\bm{\left({\overline{{\ell}^{}_{\alpha \text{L}}}}\sigma^I\gamma^\mu {\ell}^{}_{\beta\text{L}}\right)\left({\overline{Q_{\gamma \text{L}}}}\sigma^I\gamma^\mu {Q}^{}_{\delta\text{L}}\right)}$
        & \cellcolor{gray!65} $O_{ed}^{\alpha \beta \gamma \delta}$& \cellcolor{gray!65}$ {\left({\overline{E_{\alpha\text{R}}}}\gamma^\mu E_{\beta\text{R}}\right)\left({\overline{D_{\gamma\text{R}}}}\gamma^\mu D_{\delta\text{R}}\right)}$& $\bm{O_{qu}^{(1)\alpha \beta \gamma \delta}}$
        &$\bm{\left({\overline{Q_{\alpha \text{L}}}}\gamma^\mu {Q}^{}_{\beta\text{L}}\right)\left({\overline{{U}^{}_{\gamma \text{R}}}}\gamma^\mu {U}^{}_{\delta\text{R}}\right)}$\\
        & &\cellcolor{gray!65} $O_{ud}^{(1)\alpha \beta \gamma \delta}$ &\cellcolor{gray!65}{$ {\left({\overline{{U}^{}_{\alpha \text{R}}}}\gamma^\mu {U}^{}_{\beta\text{R}}\right)
        \left({\overline{D_{\gamma\text{R}}}}\gamma^\mu D_{\delta\text{R}}\right)}$}
        &\cellcolor{gray!30}{$\bm{O_{qu}^{(8)\alpha \beta \gamma \delta}}$}& \cellcolor{gray!30}{$\bm{\left({\overline{Q_{\alpha \text{L}}}}\gamma^\mu T^A{Q}^{}_{\beta\text{L}}\right)
        \left({\overline{{U}^{}_{\gamma \text{R}}}}\gamma^\mu T^A{U}^{}_{\delta\text{R}}\right)}$}\\
        & &$O_{ud}^{(8)\alpha \beta \gamma \delta}$&$\left({\overline{{U}^{}_{\alpha \text{R}}}}\gamma^\mu T^A{U}^{}_{\beta\text{R}}\right)
        \left({\overline{D_{\gamma\text{R}}}}\gamma^\mu T^AD_{\delta\text{R}}\right)$
        &\cellcolor{gray!30}{$\bm{O_{qd}^{(1)\alpha \beta \gamma \delta}}$}
        &\cellcolor{gray!30}{$\bm{\left({\overline{Q_{\alpha \text{L}}}}\gamma^\mu {Q}^{}_{\beta\text{L}}\right)\left({\overline{D_{\gamma\text{R}}}}\gamma^\mu D_{\delta}\right)}$}\\
        & & & &\cellcolor{gray!30}{$\bm{O_{qd}^{(8)\alpha \beta \gamma \delta}}$}&\cellcolor{gray!30}{$\bm{ \left({\overline{Q_{\alpha \text{L}}}}\gamma^\mu T^A{Q}^{}_{\beta\text{L}}\right)
        \left({\overline{D_{\gamma\text{R}}}}\gamma^\mu T^AD_{\delta\text{R}}\right)}$}\\
        \hline \hline \multicolumn{2}{||c||}
        {$\left(\overline{{\text{L}}^{}}\text{R}\right)\left(\overline{\text{R}} \text{L} \right) \text{and} \left(\overline{{\text{L}}^{}}\text{R}\right)\left(\overline{{\text{L}}^{}}\text{R}\right)$}
        & \multicolumn{4}{c||} {$B\text{-violating}$}\\
        \hline \cellcolor{gray!30}{$\bm{O_{\ell edq}^{\alpha \beta \gamma \delta}}$} & \cellcolor{gray!30}{$\bm{\left({\overline{{\ell}^{}_{\alpha \text{L}}}}^jE_{\beta\text{R}}\right)
        \left({\overline{D_{\gamma\text{R}}}}{Q}_{\delta\text{L}}^j\right)}$}
        & $O_{duq}^{\alpha \beta \gamma}$ & \multicolumn{3}{c||} {$\epsilon^{abc} \epsilon_{j k}\left[\left(D_{\alpha\text{R}}^{a}
        \right)^{\text{T}} {\sf C} {U}_{\beta\text{R}}^{b}\right]\left[\left({Q}_{\gamma\text{L}}^{c j}\right)^{\text{T}} {\sf C} {\ell}_{\delta\text{L}}^{k}\right] $}  \\
        \cellcolor{gray!30}{$\bm{O_{quqd}^{(1)\alpha \beta \gamma \delta}}$}& \cellcolor{gray!30}{$\bm{\left({\overline{Q_{\alpha \text{L}}}}^j{U}^{}_{\beta\text{R}}\right)\epsilon_{jk}
        \left({\overline{Q}}_{\gamma}^kD_{\delta\text{R}}\right)}$}
        & $O_{qqu}^{\alpha \beta \gamma}$ & \multicolumn{3}{c||} {$\epsilon^{abc} \epsilon_{j k}\left[\left({Q}_{\alpha\text{L}}^{a j}
        \right)^{\text{T}} {\sf C} {Q}_{\beta\text{L}}^{b k}\right]\left[\left({U}_{\gamma\text{R}}^{c}\right)^{\text{T}} {\sf C} E_{\delta \text{R}}\right] $}  \\
        $O_{quqd}^{(8)\alpha \beta \gamma \delta}$ & $\left({\overline{Q_{\alpha}}}^j T^A {U}^{}_{\beta\text{R}}\right)\epsilon_{jk}\left({\overline{Q}}_{\gamma}^k T^A D_{\delta\text{R}}\right)$
        & $O_{qdd}^{\alpha \beta \gamma}$ & \multicolumn{3}{c||} {$\epsilon^{abc} \epsilon_{j n} \epsilon_{k m}
        \left[\left({Q}_{\alpha \text{L}}^{a j}\right)^{\text{T}} {\sf C} {Q}_{\beta\text{L}}^{b k}\right]\left[\left({Q}_{\gamma\text{L}}^{c m}\right)^{\text{T}} {\sf C} {\ell}_{\delta\text{L}}^{n}\right]$} \\
        \cellcolor{gray!30}{$\bm{O_{\ell equ}^{(1)\alpha \beta \gamma \delta}}$} & \cellcolor{gray!30}{$\bm{\left({\overline{{\ell}^{}_{\alpha \text{L}}}}^jE_{\beta \text{R}}\right)
        \epsilon_{jk}\left({\overline{Q_{\gamma \text{L}}}}^k{U}^{}_{\delta\text{R}}\right)}$}
         & $O_{duu}^{\alpha \beta \gamma}$ & \multicolumn{3}{c||} {$\epsilon^{abc}\left[\left(D_{\alpha\text{R}}^{a}\right)^{\text{T}}
        {\sf C} {U}_{\beta\text{R}}^{b}\right]\left[\left({U}_{\gamma\text{R}}^{c}\right)^{\text{T}}{\sf C}E_{\delta\text{R}}\right]$}\\
        $O_{\ell equ}^{(3)\alpha \beta \gamma \delta} $& $ \left({\overline{{\ell}^{}_{\alpha \text{L}}}}^j \sigma_{\mu\nu} E_{\beta\text{R}}\right)\epsilon_{jk}
        \left({\overline{Q_{\gamma \text{L}}}}^k \sigma^{\mu\nu}{U}^{}_{\delta\text{R}}\right)$ &
        & \multicolumn{3}{c||} { } \\
        \hline \hline
    \end{tabular}
  }
\end{center}
\vspace{0.4cm}
\caption{Summary of the dimension-six operators in the Warsaw basis of the SMEFT~\cite{Grzadkowski:2010es}. There are 33 operators that are induced by integrating out heavy fermionic triplets of the type-III seesaw model at one-loop level, and they are highlighted in boldface. For comparison, we show the dim-6 operators in the SEFT-I in light gray, while those in the SEFT-II in dark gray and light gray. It can be observed that the SEFT-II covers all the dim-6 operators in the SEFT-I and SEFT-III, whereas the SEFT-III contains two more dim-6 operators than the SEFT-I does, namely, the operators $O^{}_W$ and $O_{qq}^{(3)}$. }
\label{tab:warsawbasis}
\end{table}

\begin{itemize}
	\item $X^3$
	\begin{eqnarray}
	C^{}_{W} &=& -\frac{g_2^3}{1440\pi^2} \tr\left(M_\Sigma^{-2}\right) \;.
	\label{eq:3W}
	\end{eqnarray}
	\item $X^2H^2$
	\begin{eqnarray}
	C^{}_{HB} &=& \frac{g_1^2}{128\pi^2}\tr\left(Y_\Sigma^{} M_\Sigma^{-2}Y_\Sigma^\dag\right) \;,
	\label{eq:HBw}
	\\
	C^{}_{HWB} &=& -\frac{g^{}_1g^{}_2}{64\pi^2}\tr\left(Y_\Sigma^{}M_\Sigma^{-2}Y_\Sigma^\dag\right) \;,
	\label{eq:HWBw}
	\\
	C^{}_{HW} &=&  \frac{7g_2^2}{384\pi^2}\tr\left(Y_\Sigma^{}M_\Sigma^{-2}Y_\Sigma^\dag\right) \;.
	\label{eq:HWw}
	\end{eqnarray}
	
	\item $H^4D^2$
	\begin{eqnarray}
	C^{}_{H\Box} &=& \frac{g_2^2}{128\pi^2}\tr\left[Y^{}_\Sigma \frac{\left(7+2L^{}_\Sigma\right)}{M_\Sigma^2}Y_\Sigma^\dag\right] -\frac{g_2^4}{160\pi^2}\tr\left(M_\Sigma^{-2}\right)-\frac{g_1^2}{384\pi^2} \tr\left[Y^{}_\Sigma\frac{\left(5+6L^{}_\Sigma\right)}{M_\Sigma^2}Y_\Sigma^\dag\right] \nonumber \\
	&&
	-\frac{\big(Y_\Sigma^\dag Y_\Sigma^{}\big)^{}_{ki} \big(Y_\Sigma^\dag Y_\Sigma^{}\big)^{}_{ik} L^{}_{ik}}{16\pi^2\left(M_i^2-M_k^2\right)} + \frac{\big(Y_\Sigma^\dag Y_\Sigma^{}\big)^{}_{ik} \big(Y_\Sigma^\dag Y_\Sigma^{}\big)^{}_{ki}}{32\pi^2\left(M_i^2-M_k^2\right)^3} \bigg[M_i^4\left(1-L^{}_{ik}\right) - M_k^4\left(1+L^{}_{ik}\right)\bigg] \nonumber \\
	&& -\frac{\big(Y_\Sigma^\dag Y_\Sigma^{}\big)^{}_{ik} \big(Y_\Sigma^\dag Y_\Sigma^{}\big)^{}_{ik}}{32\pi^2 M^{}_i M^{}_k \left(M_i^2-M_k^2\right)^3} \bigg[ M_i^6\left(1+L^{}_k\right) -M_i^4 M_k^2\left(13+6L^{}_i-3L^{}_k\right)
	  \nonumber \\
	&& -M_k^6\left(1+L^{}_i\right)  + M_k^4M_i^2 \left(13+6L^{}_k - 3L^{}_i\right)\bigg] -\frac{1}{64\pi^2} \tr\left(Y_l^{}Y_l^\dag Y^{}_\Sigma \frac{\left(11+6L^{}_\Sigma\right)}{M_\Sigma^2} Y_\Sigma^\dag\right) \;, \qquad
	\label{eq:HBoxw}
	\\
	C^{}_{HD} &=& -\frac{g_1^2}{96\pi^2}\tr\left[Y^{}_\Sigma \frac{\left(5+6L^{}_\Sigma\right)}{M_\Sigma^2} Y_\Sigma^\dag\right]  -\frac{L^{}_{ik} }{32\pi^2\left(M_i^2-M_k^2\right)} \big(Y_\Sigma^\dag Y_\Sigma^{}\big)^{}_{ki} \big(Y_\Sigma^\dag Y_\Sigma^{}\big)^{}_{ik}\nonumber\\
	&&
	-\frac{\big(Y_\Sigma^\dag Y_\Sigma^{}\big)^{}_{ik} \big(Y_\Sigma^\dag Y_\Sigma^{}\big)^{}_{ik}}{16\pi^2M^{}_iM^{}_k \left(M_i^2-M_k^2\right)^3} \bigg[M_i^6 \left(1+L^{}_k\right) - M_k^6\left(1+L^{}_i\right) + M_i^4 M_k^2\left(5+3L^{}_i - 6L^{}_k\right)  \nonumber \\
	&&  -M_k^4 M_i^2 \left(5+3L^{}_k-6L^{}_i\right)\bigg]
	+\frac{\big(Y_\Sigma^\dag Y_\Sigma^{}\big)^{}_{ik} \big(Y_\Sigma^\dag Y_\Sigma^{}\big)^{}_{ki}}{4\pi^2\left(M_i^2-M_k^2\right)^3} \bigg[M_i^4\left(1-L^{}_{ik}\right) - M_k^4\left(1+L^{}_{ik}\right)\bigg] \nonumber \\
	&& -\frac{1}{32\pi^2}\tr\left[Y_l^{} Y_l^\dag Y^{}_\Sigma \frac{\left(11+6L^{}_\Sigma\right)}{M_\Sigma^2}Y_\Sigma^\dag\right] \;.
	\label{eq:HDw}
	\end{eqnarray}
	
	\item $H^6$
	\begin{eqnarray}
	C^{}_H &=& \frac{\lambda g_2^2}{48\pi^2} \tr \left[Y_\Sigma^{}\frac{\left(7+2L_\Sigma\right)}{M_\Sigma^2}Y_\Sigma^\dag\right]-\frac{\lambda g_2^4}{60\pi^2}\tr\left(M_\Sigma^{-2}\right)
	\nonumber \\
	&& -\frac{\lambda\ \tr\left[Y_l^{}Y_l^\dag Y_\Sigma^{}\frac{\left(7-2L_\Sigma\right)}{M_\Sigma^2}Y_\Sigma^\dag\right]}{16\pi^2}  +\frac{\lambda\big(Y_\Sigma^\dag Y_\Sigma^{}\big)_{ik}\big(Y_\Sigma^\dag Y_\Sigma^{}\big)_{ik}}{16\pi^2M_iM_k}  +\frac{\lambda^2}{4\pi^2}\tr\left(Y_\Sigma^{}M_\Sigma^{-2}Y_\Sigma^\dag\right) \nonumber \\
	&&
	-\frac{\tr\left[Y_l^{}Y_l^\dag Y_\Sigma^{}\frac{\left(1+L_\Sigma\right)}{M_\Sigma^2}Y_\Sigma^\dag Y_l^{}Y_l^\dag\right]}{4\pi^2}-\frac{7\lambda}{8\pi^2}\frac{\big(Y_\Sigma^\dag Y_\Sigma^{}\big)_{ki}\big(Y_\Sigma^\dag Y_\Sigma^{}\big)_{ik}L_{ik}}{\left(M_i^2-M_k^2\right)} \nonumber \\
	&&
	-\frac{\lambda\big(Y_\Sigma^\dag Y_\Sigma^{}\big)_{ik}\big(Y_\Sigma^\dag Y_\Sigma^{}\big)_{ki}}{2 \pi^2\left(M_i^2-M_k^2\right)^3}\bigg[M_i^4\left(1-L_{ik}\right)-M_k^4\left(1+L_{ik}\right)\bigg] \nonumber \\
	&&
	-\frac{\lambda\big(Y_\Sigma^\dag Y_\Sigma^{}\big)_{ik}\big(Y_\Sigma^\dag Y_\Sigma^{}\big)_{ik}}{8\pi^2M_iM_k\left(M_i^2-M_k^2\right)^3}\bigg[M_i^6\left(1+L_k\right)-M_k^6\left(1+L_i\right)   \nonumber \\
	&&   -M_i^4M_k^2\left(11+5L_i-2L_k\right)+M_k^4M_i^2\left(11+5L_k-2L_i\right)\bigg] \nonumber \\
	&&
	+\frac{\big(Y_\Sigma^\dag Y_\Sigma^{}\big)_{ij}\big(Y_\Sigma^\dag Y_\Sigma^{}\big)_{jk}\big(Y_\Sigma^\dag Y_\Sigma^{}\big)_{ik}}{8\pi^2}\frac{M_iM_k\big(M_i^2L_{jk}+M_j^2L_{ki}+M_k^2L_{ij}\big)}{\big(M_i^2-M_j^2 \big)\big(M_j^2-M_k^2\big)\left(M_k^2-M_i^2\right)}\nonumber\\
    &&
	-\frac{3\big(Y_\Sigma^\dag Y_\Sigma^{}\big)_{ij}\big(Y_\Sigma^\dag Y_\Sigma^{}\big)_{jk}\big(Y_\Sigma^\dag Y_\Sigma^{}\big)_{ki}}{8\pi^2}\frac{\big(M_i^2M_j^2L_{ij}+M_j^2M_k^2L_{jk}+M_i^2M_k^2L_{ki}\big)}{\big(M_i^2-M_j^2\big)\big(M_j^2-M_k^2\big)\left(M_k^2-M_i^2\right)}\nonumber \\
	&& +\frac{\left(Y_\Sigma^\dag Y_l^{}Y_l^\dag Y_\Sigma^{}\right)_{ki}\left(Y_\Sigma^\dag Y_\Sigma^{}\right)_{ik}L_{ik}}{2\pi^2\left(M_i^2-M_k^2\right)}\;.
	\label{eq:H6}
	\end{eqnarray}
	
	\item $\psi^2 X H$
	\begin{eqnarray}
	C^{\alpha\beta}_{eB} &=& \frac{g_1}{128\pi^2}\left(Y_\Sigma^{}M_\Sigma^{-2}Y_\Sigma^\dag Y_l^{}\right)^{\alpha\beta} \;,
	\label{eq:eBw}
	\\
	C^{\alpha\beta}_{eW} &=& \frac{3g_2}{128\pi^2}\left(Y_\Sigma^{}M_\Sigma^{-2}Y_\Sigma^\dag Y_l^{}\right)^{\alpha\beta} \;.
	\label{eq:eWw}
	\end{eqnarray}
	
	\item $\psi^2 H^2D$
	\begin{eqnarray}
	C^{(1)\alpha\beta}_{Hq} &=& \frac{-g_1^2}{1152\pi^2}\delta^{\alpha\beta}\tr\left[Y_\Sigma^{}\frac{\left(5+6L_\Sigma\right)}{M_\Sigma^2}Y_\Sigma^\dag\right] \;,
	\label{eq:HQ1w}
	\\
	C^{(3)\alpha\beta}_{Hq} &=& \frac{-g_2^4}{240\pi^2}\delta^{\alpha\beta}\tr (M_\Sigma^{-2})+\frac{g_2^2}{384\pi^2}\delta^{\alpha\beta}\tr\left[Y_\Sigma^{}\frac{\left(7+2L_\Sigma\right)}{M_\Sigma^2}Y_\Sigma^\dag\right] \;,
	\label{eq:HQ3w}
	\\
	C^{\alpha\beta}_{Hu} &=& \frac{-g_1^2}{288\pi^2}\delta^{\alpha\beta}\tr\left[Y_\Sigma^{}\frac{\left(5+6L_\Sigma\right)}{M_\Sigma^2}Y_\Sigma^\dag\right] \;,
	\label{eq:HUw}
	\\
	C^{\alpha\beta}_{Hd} &=& \frac{g_1^2}{576\pi^2}\delta^{\alpha\beta}\tr\left[Y_\Sigma^{}\frac{\left(5+6L_\Sigma\right)}{M_\Sigma^2}Y_\Sigma^\dag\right] \;,
	\label{eq:HDDw}
	\end{eqnarray}
	\vspace{-0.8cm}
	\begin{eqnarray}
	C^{(1)\alpha\beta}_{H\ell} &=&\frac{g_1^2}{1536\pi^2}\left\{4\delta^{\alpha\beta}\tr\left[Y_\mathrm{\Sigma}\frac{\left(5+6L_\mathrm{\Sigma}\right)}{M_\mathrm{\Sigma}^2}Y_\mathrm{\Sigma}^\dag\right]+11\left[Y_\mathrm{\Sigma}\frac{\left(11+6L_\mathrm{\Sigma}\right)}{M_\mathrm{\Sigma}^2}Y_\mathrm{\Sigma}^\dag\right]^{\alpha\beta}\right\} \nonumber \\
	&& -\frac{3g_2^2}{512\pi^2}\left[Y_\mathrm{\Sigma}\frac{\left(7+30L_\mathrm{\Sigma}\right)}{M_\mathrm{\Sigma}^2}Y_\mathrm{\Sigma}^\dag\right]^{\alpha\beta} -\frac{3L_{ik}}{32\pi^2}\frac{\left(Y_\Sigma^\dag Y_\Sigma^{}\right)_{ik}\left(Y_\Sigma^\dag\right)_{k\beta}\left(Y_\Sigma^{}\right)_{\alpha i}}{\left(M_i^2-M_k^2\right)}\nonumber \\
	&& +\frac{3}{256\pi^2}\left[Y_l^{}Y_l^\dag Y_\Sigma^{}\frac{\left(3+2L_\Sigma\right)}{M_\Sigma^2}Y_\Sigma^\dag + {\rm h.c.}\right]^{\alpha\beta}\nonumber \\
	&& -\frac{9}{128\pi^2}\tr\left[Y_\Sigma^\dag Y_\Sigma^{}(1+2L_\Sigma^{})\right]\left(Y_\Sigma^{} M_\Sigma^{-2}Y_\Sigma^\dag\right)_{\alpha \beta}  \nonumber \\
	&& -\frac{9\left(Y_\Sigma^\dag Y_\Sigma^{}\right)_{ik}\left(Y_\Sigma^\dag\right)_{k\beta}\left(Y_\Sigma^{}\right)_{\alpha i}}{512\pi^2M_i^2M_k^2}\bigg[M_i^2\left(3+2L_i\right)+M_k^2\left(3+2L_k\right)\bigg] \nonumber \\
	&& -\frac{3L_{ik}\left(M_i^2+M_k^2\right)}{128\pi^2M_iM_k\left(M_i^2-M_k^2\right)}\left(Y_\Sigma^\dag Y_\Sigma^{}\right)_{ik}\left(Y_\Sigma^\dag\right)_{i\beta}\left(Y_\Sigma^{}\right)_{\alpha k}\;,
	\label{eq:Hl1w}
	\\
	C^{(3)\alpha\beta}_{H\ell} &=& +\frac{g_2^2}{1536\pi^2}\left\{4\delta^{\alpha\beta}\tr\left[Y_\mathrm{\Sigma}\frac{\left(7+2L_\mathrm{\Sigma}\right)}{M_\mathrm{\Sigma}^2}Y_\mathrm{\Sigma}^\dag\right]-\left[Y_\mathrm{\Sigma}\frac{\left(127+158L_\mathrm{\Sigma}\right)}{M_\mathrm{\Sigma}^2}Y_\mathrm{\Sigma}^\dag\right]^{\alpha\beta}\right\} \nonumber \\
	&& -\frac{g_2^4\delta^{\alpha\beta}}{240\pi^2}\tr\left(  M_\mathrm{\Sigma}^{-2}\right) + \frac{g_1^2}{512\pi^2}\left[Y_\mathrm{\Sigma}\frac{\left(11+6L_\mathrm{\Sigma}\right)}{M_\mathrm{\Sigma}^2}Y_\mathrm{\Sigma}^\dag\right]^{\alpha\beta} \nonumber \\
	&&  -\frac{3}{128\pi^2}\tr\left[Y_\Sigma^\dag Y_\Sigma^{}(1+2L_\Sigma^{})\right]\left(Y_\Sigma^{} M_\Sigma^{-2}Y_\Sigma^\dag\right)_{\alpha\beta}  \nonumber \\
	&&  +\frac{3}{256\pi^2}\left[Y_l^{}Y_l^\dag Y_\Sigma^{}\frac{\left(3+2L_\Sigma\right)}{M_\Sigma^2}Y_\Sigma^\dag+{\rm h.c.}\right]^{\alpha\beta} -\frac{L_{ik}}{16\pi^2}\frac{\left(Y_\Sigma^\dag Y_\Sigma^{}\right)_{ik}\left(Y_\Sigma^\dag\right)_{k\beta}\left(Y_\Sigma^{}\right)_{\alpha i}}{\left(M_i^2-M_k^2\right)}
	 \nonumber \\
	&& 	-\frac{\left(Y_\Sigma^\dag Y_\Sigma^{}\right)_{ik}\left(Y_\Sigma^\dag\right)_{i\beta}\left(Y_\Sigma^{}\right)_{\alpha k}}{128\pi^2M_iM_k}\left(1-L_i-L_k\right)
    \nonumber \\
	&&	+\frac{\left(Y_\Sigma^\dag Y_\Sigma^{}\right)_{ik}\left(Y_\Sigma^\dag\right)_{i\beta}\left(Y_\Sigma^{}\right)_{\alpha k}}{32\pi^2M_iM_k\left(M_i^2-M_k^2\right)}\bigg[M_i^2\left(1+L_k\right)-M_k^2\left(1+L_i\right)\bigg]\nonumber \\
	&& -\frac{3\left(Y_\Sigma^\dag Y_\Sigma^{}\right)_{ik}\left(Y_\Sigma^\dag\right)_{k\beta}\left(Y_\Sigma^{}\right)_{\alpha i}}{512\pi^2M_i^2M_k^2}\bigg[M_i^2\left(3+2L_i\right)+M_k^2\left(3+2L_k\right)\bigg]
	 \;,
	 \label{eq:Hl3w}
	 \\
	C^{\alpha\beta}_{He} &=& \frac{g_1^2}{192\pi^2}\delta^{\alpha\beta}\tr\left[Y_\Sigma^{}\frac{\left(5+6L_\Sigma^{}\right)}{M_\Sigma^2}Y_\Sigma^\dag\right]+\frac{1}{128\pi^2}\left[Y_l^\dag Y_\Sigma^{}\frac{\left(19-6L_\Sigma^{}\right)}{M_\Sigma^2}Y_\Sigma^\dag Y_l^{}\right]^{\alpha\beta} \;.
	\label{eq:HEw}
	\end{eqnarray}
	
	\item $\psi^2H^3$
	\begin{eqnarray}
	C^{\alpha\beta}_{uH} &=& \frac{-g_2^4}{240\pi^2}Y_{\rm u}^{\alpha\beta}\tr\left(M_\Sigma^{-2}\right)+\frac{g_2^2}{192\pi^2}Y_{\rm u}^{\alpha\beta}\tr\left[Y_\Sigma^{}\frac{\left(7+2L_\Sigma\right)}{M_\Sigma^2}Y_\Sigma^\dag\right] \nonumber \\
	&& -\frac{Y_{\rm u}^{\alpha\beta}}{64\pi^2}\ \tr\left[Y_l^{}Y_l^\dag Y_\Sigma^{}\frac{\left(7-2L_\Sigma\right)}{M_\Sigma^2}Y_\Sigma^\dag\right] +\frac{\lambda Y_{\rm u}^{\alpha\beta}}{8\pi^2}\tr\left(Y_\Sigma^{}M_\Sigma^{-2}Y_\Sigma^\dag\right)
	\nonumber \\
	&& - \frac{Y_{\rm u}^{\alpha\beta}\left(Y_\Sigma^\dag Y_\Sigma^{}\right)_{ik}\left(Y_\Sigma^\dag Y_\Sigma^{}\right)_{ik}}{32\pi^2M_iM_k\left(M_i^2-M_k^2\right)^3}\bigg[M_i^6\left(1+L_k\right)-M_k^6L_i   -M_i^4M_k^2\left(10+4L_i-L_k\right) \nonumber \\
	&& +3M_k^4M_i^2\left(3+2L_k-L_i\right)\bigg] -\frac{7}{32\pi^2}\frac{Y_{\rm u}^{\alpha\beta}\left(Y_\Sigma^\dag Y_\Sigma^{}\right)_{ik}\left(Y_\Sigma^\dag Y_\Sigma^{}\right)_{ki}L_{ik}}{M_i^2-M_k^2} \nonumber \\
	&& -\frac{Y_{\rm u}^{\alpha\beta}\left(Y_\Sigma^\dag Y_\Sigma^{}\right)_{ik}\left(Y_\Sigma^\dag Y_\Sigma^{}\right)_{ki}}{8\pi^2\left(M_i^2-M_k^2\right)^3}\bigg[M_i^4\left(1-L_{ik}\right)-M_k^4\left(1+L_{ik}\right)\bigg] \;,
	\label{eq:UHw}
	\\
	C^{\alpha\beta}_{dH} &=& \frac{-g_2^4}{240\pi^2}Y_{\rm d}^{\alpha\beta}\tr\left(M_\Sigma^{-2}\right)+\frac{g_2^2}{192\pi^2}Y_{\rm d}^{\alpha\beta}\tr\left[Y_\Sigma^{}\frac{\left(7+2L_\Sigma\right)}{M_\Sigma^2}Y_\Sigma^\dag\right] \nonumber \\
	&& -\frac{Y_{\rm d}^{\alpha\beta}}{64\pi^2}\ \tr\left[Y_l^{}Y_l^\dag Y_\Sigma^{}\frac{\left(7-2L_\Sigma\right)}{M_\Sigma^2}Y_\Sigma^\dag\right]
	+\frac{\lambda Y_{\rm d}^{\alpha\beta}}{8\pi^2}\tr\left(Y_\Sigma^{}M_\Sigma^{-2}Y_\Sigma^\dag\right) \nonumber \\
	&& -\frac{Y_{\rm d}^{\alpha\beta}\left(Y_\Sigma^\dag Y_\Sigma^{}\right)_{ik}\left(Y_\Sigma^\dag Y_\Sigma^{}\right)_{ik}}{32\pi^2M_iM_k\left(M_i^2-M_k^2\right)^3}\bigg[M_i^6\left(1+L_k\right)-M_k^6L_i -M_i^4M_k^2\left(10+4L_i-L_k\right) \nonumber \\
	&&  +3M_k^4M_i^2\left(3+2L_k-L_i\right)\bigg] -\frac{7}{32\pi^2}\frac{Y_{\rm d}^{\alpha\beta}\left(Y_\Sigma^\dag Y_\Sigma^{}\right)_{ik}\left(Y_\Sigma^\dag Y_\Sigma^{}\right)_{ki}L_{ik}}{M_i^2-M_k^2} \nonumber \\
	&&
	-\frac{Y_{\rm d}^{\alpha\beta}\left(Y_\Sigma^\dag Y_\Sigma^{}\right)_{ik}\left(Y_\Sigma^\dag Y_\Sigma^{}\right)_{ki}}{8\pi^2\left(M_i^2-M_k^2\right)^3}\bigg[M_i^4\left(1-L_{ik}\right)-M_k^4\left(1+L_{ik}\right)\bigg] \;,
	\label{eq:DHw}
	\\
	C^{\alpha\beta}_{eH} &=& -\frac{g_2^4}{240\pi^2}Y_l^{\alpha\beta}\tr\left(M_\Sigma^{-2}\right) -\frac{g_1^2}{16\pi^2}\left[Y_\Sigma^{}\frac{\left(1+3L_\Sigma\right)}{M_\Sigma^2}Y_\Sigma^\dag Y_l^{}\right]^{\alpha\beta} \nonumber \\
	&& +\frac{g_2^2}{192\pi^2}\left\{Y_l^{\alpha\beta}\tr\left[Y_\Sigma^{}\frac{\left(7+2L_\Sigma\right)}{M_\Sigma^2}Y_\Sigma^\dag\right]-24\left[Y_\Sigma^{}\frac{\left(1+3L_\Sigma\right)}{M_\Sigma^2}Y_\Sigma^\dag Y_l^{}\right]^{\alpha\beta}\right\} \nonumber \\
	&& +\frac{\lambda}{32\pi^2}\left[Y_\Sigma^{}\frac{\left(41+30L_\Sigma\right)}{M_\Sigma^2}Y_\Sigma^\dag Y_l^{}\right]^{\alpha\beta}+\frac{\lambda Y_l^{\alpha\beta}}{8\pi^2}\tr\left(Y_\Sigma^{}M_\Sigma^{-2}Y_\Sigma^\dag\right) \nonumber \\
	&& +\frac{1}{128\pi^2}\left[Y_\Sigma^{}\frac{\left(7+10L_\Sigma\right)}{M_\Sigma^2}Y_\Sigma^\dag Y_l^{}Y_l^\dag Y_l^{}\right]^{\alpha\beta} +\frac{1}{64\pi^2}\left[Y_l^{}Y_l^\dag Y_\Sigma^{}\frac{\left(17+6L_\Sigma\right)}{M_\Sigma^2}Y_\Sigma^\dag Y_l^{}\right]^{\alpha\beta}
	\nonumber \\
	&& -\frac{{9\left(Y_\Sigma^\dag Y_\Sigma^{}\right)}_{kk}\left(Y_\Sigma^{}\right)_{\alpha i}\left(Y_\Sigma^\dag Y_l^{}\right)_{i\beta}\left(1+2L_k\right)}{64\pi^2M_i^2}
	-\frac{Y_l^{\alpha\beta}}{64\pi^2}\tr\left[Y_l^{}Y_l^\dag Y_\Sigma^{}\frac{\left(7-2L_\Sigma\right)}{M_\Sigma^2}Y_\Sigma^\dag\right] \nonumber \\
	&& +\frac{\left(Y_\Sigma^\dag Y_\Sigma^{}\right)_{ik}\left(Y_\Sigma^{}\right)_{\alpha k}\left(Y_\Sigma^\dag Y_l^{}\right)_{i\beta}}{128\pi^2M_iM_k\left(M_i^2-M_k^2\right)^2}\bigg[\left(19+20L_i-6L_k\right)M_i^4+\left(15+10L_i+4L_k\right)M_k^4   \nonumber \\
	&&  -2\left(17+13L_i+L_k\right)M_i^2M_k^2\bigg] -\frac{\left(Y_\Sigma^\dag Y_\Sigma^{}\right)_{ki}\left(Y_\Sigma^{}\right)_{\alpha k}\left(Y_\Sigma^\dag Y_l^{}\right)_{i\beta}}{128\pi^2M_i^2M_k^2\left(M_i^2-M_k^2\right)^2}\bigg[6\left(7+6L_i\right)M_i^6  \nonumber \\
	&& +3\left(3+2L_k\right)M_k^6-\left(95+132L_i-66L_k\right)M_i^4M_k^2  +4\left(11+19L_i-13L_k\right)M_i^2M_k^4\bigg] \nonumber\\
	&&-\frac{Y_l^{\alpha\beta}\left(Y_\Sigma^\dag Y_\Sigma^{}\right)_{ik}\left(Y_\Sigma^\dag Y_\Sigma^{}\right)_{ik}}{32\pi^2M_iM_k\left(M_i^2-M_k^2\right)^3}\bigg[L_kM_i^6  -\left(1+L_i\right)M_k^6-3\left(3+2L_i-L_k\right)M_i^4M_k^2 \nonumber \\
	&&  +\left(10-L_i+4L_k\right)M_i^2M_k^4\bigg] -\frac{Y_l^{\alpha\beta}\left(Y_\Sigma^\dag Y_\Sigma^{}\right)_{ik}\left(Y_\Sigma^\dag Y_\Sigma^{}\right)_{ki}}{32\pi^2\left(M_i^2-M_k^2\right)^3}\bigg[\left(4+3L_{ik}\right)M_i^4 \nonumber \\
	&& -\left(4-3L_{ik}\right)M_k^4-14L_{ik}M_i^2M_k^2\bigg] \;.
	\label{eq:eHw}
	\end{eqnarray}
	
	\item Four-quark
	\begin{eqnarray}
	C^{(3)\alpha\beta\gamma\delta}_{qq} &=& -\frac{g_2^4}{480\pi^2}\delta^{\alpha\beta}\delta^{\gamma\delta}\ \tr\left(M_\Sigma^{-2}\right) \;,
	\label{eq:qq3}
	\\
	C^{(1)\alpha\beta\gamma\delta}_{qu} &=& -\frac{1}{96\pi^2}{Y_{\rm u}^\dag}^{\gamma\beta}Y_{\rm u}^{\alpha\delta}\ \tr\left(Y_\Sigma^{}M_\Sigma^{-2}Y_\Sigma^\dag\right) \;,
	\label{eq:QU1w}
	\\
	C^{(8)\alpha\beta\gamma\delta}_{qu} &=& -\frac{1}{16\pi^2}{Y_{\rm u}^\dag}^{\gamma\beta}Y_{\rm u}^{\alpha\delta}\ \tr\left(Y_\Sigma^{}M_\Sigma^{-2}Y_\Sigma^\dag\right) \;,
	\label{eq:QU8w}
	\\
	C^{(1)\alpha\beta\gamma\delta}_{qd} &=& -\frac{1}{96\pi^2}{Y_{\rm d}^\dag}^{\gamma\beta}Y_{\rm d}^{\alpha\delta}\ \tr\left(Y_\Sigma^{}M_\Sigma^{-2}Y_\Sigma^\dag\right) \;,
	\label{eq:Qd1w}
	\\
	C^{(8)\alpha\beta\gamma\delta}_{qd} &=& -\frac{1}{16\pi^2}{Y_{\rm d}^\dag}^{\gamma\beta}Y_{\rm d}^{\alpha\delta}\ \tr\left(Y_\Sigma^{}M_\Sigma^{-2}Y_\Sigma^\dag\right) \;,
	\label{eq:Qd8w}
	\\
	C^{(1)\alpha\beta\gamma\delta}_{quqd} &=& \frac{1}{16\pi^2}Y_{\rm u}^{\alpha\beta}Y_{\rm d}^{\gamma\delta}\ \tr\left(Y_\Sigma^{}M_\Sigma^{-2}Y_\Sigma^\dag\right) \;.
	\label{eq:QUQdw}
	\end{eqnarray}
	
	\item Four-lepton
	\begin{eqnarray}
	C^{\alpha\beta\gamma\delta}_{\ell\ell} &=& \frac{-g_2^4}{480\pi^2}\left(2\delta^{\alpha\delta}\delta^{\beta\gamma}-\delta^{\alpha\beta}\delta^{\gamma\delta}\right)\tr\left(M_\Sigma^{-2}\right)-\frac{g_1^2}{768\pi^2}\delta^{\alpha\beta}\left[Y_\Sigma^{}\frac{\left(11+6L_\Sigma\right)}{M_\Sigma^2}Y_\Sigma^\dag\right]^{\gamma\delta}\nonumber \\
	&& + \frac{g_2^2}{768\pi^2}\left\{-\delta^{\gamma\delta}\left[Y_\Sigma^{}\frac{\left(13+2L_\Sigma\right)}{M_\Sigma^2}Y_\Sigma^\dag\right]^{\alpha\beta}+2\delta^{\beta\gamma }\left[Y_\Sigma^{}\frac{\left(13+2L_\Sigma\right)}{M_\Sigma^2}Y_\Sigma^\dag\right]^{\alpha\delta}\right\}\nonumber \\
	&&-\frac{1}{128\pi^2}\left\{ \left[Y_\Sigma^{}\frac{\left(3+2L_\Sigma\right)}{M_\Sigma^2}Y_\Sigma^\dag\right]^{\alpha\delta}\left(Y_l^{}Y_l^\dag\right)^{\gamma\beta}+(\beta \leftrightarrow \delta)\right\}\nonumber \\
	&&  -\frac{\left(Y_\Sigma^{}\right)^{\alpha i}\left(Y_\Sigma^{}\right)^{\gamma k}\left(Y_\Sigma^\dag\right)^{i \delta}\left(Y_\Sigma^\dag\right)^{k\beta}L_{ik}}{128\pi^2\left(M_i^2-M_k^2\right)} -\frac{\left(Y_\Sigma^{}\right)^{\alpha i}\left(Y_\Sigma^{}\right)^{\gamma k}\left(Y_\Sigma^\dag\right)^{k \delta}\left(Y_\Sigma^\dag\right)^{i\beta}L_{ik}}{32\pi^2\left(M_i^2-M_k^2\right)}\nonumber\\
	&& -\left(Y_\Sigma^{}\right)^{\alpha k}\left(Y_\Sigma^{}\right)^{\gamma k}\left(Y_\Sigma^\dag\right)^{i \delta}\left(Y_\Sigma^\dag\right)^{i\beta}\frac{M_i^2\left(1+L_k\right)-M_k^2\left(1+L_i\right)}{64\pi^2M_iM_k\left(M_i^2-M_k^2\right)}	\;,
	\label{eq:llw}
	\\
	C^{\alpha\beta\gamma\delta}_{\ell e} &=& -\frac{g_1^2}{384\pi^2}\left[Y_\Sigma^{}\frac{\left(11+6L_\Sigma\right)}{M_\Sigma^2}Y_\Sigma^\dag\right]^{\alpha\beta}\delta^{\gamma\delta} + \frac{3}{128\pi^2}\left[Y_\Sigma^{}\frac{\left(3+2L_\Sigma\right)}{M_\Sigma^2}Y_\Sigma^\dag\right]^{\alpha\beta}\left(Y_l^\dag Y_l^{}\right)^{\gamma\delta} \nonumber \\
	&& -\frac{3}{64\pi^2}\left[Y_l^{\alpha\delta}\left(Y_l^\dag Y_\Sigma^{}M_\Sigma^{-2}Y_\Sigma^\dag\right)^{\gamma\beta}+{\rm h.c.}\right] \nonumber \\
	&&-\frac{1}{32\pi^2}Y_l^{\alpha\delta}{Y_l^\dag}^{\gamma\beta}\tr\left(Y_\Sigma^{}M_\Sigma^{-2}Y_\Sigma^\dag\right) \;.
	\label{eq:lew}
	\end{eqnarray}
	
	\item Semileptonic
	\begin{eqnarray}
	C^{(1)\alpha\beta\gamma\delta}_{\ell q} &=& \frac{g_1^2}{2304\pi^2}\left[Y_\Sigma^{}\frac{\left(11+6L_\Sigma\right)}{M_\Sigma^2}Y_\Sigma^\dag\right]^{\alpha\beta}\delta^{\gamma\delta} \nonumber \\
	&& +\frac{3}{256\pi^2}\left(Y_{\rm u}^{}Y_{\rm u}^\dag-Y_{\rm d}^{}Y_{\rm d}^\dag\right)^{\gamma\delta}\left[Y_\Sigma^{}\frac{\left(3+2L_\Sigma\right)}{M_\Sigma^2}Y_\Sigma^\dag\right]^{\alpha\beta} \;,
	\label{eq:lQ1w}
	\\
	C^{(3)\alpha\beta\gamma\delta}_{\ell q} &=& -\frac{g_2^4}{240\pi^2}\delta^{\alpha\beta}\delta^{\gamma\delta} \tr (M_\Sigma^{-2})+\frac{g_2^2}{768\pi^2}\left[Y_\Sigma^{}\frac{\left(13+2L_\Sigma\right)}{M_\Sigma^2}Y_\Sigma^\dag\right]^{\alpha\beta}\delta^{\gamma\delta} \nonumber \\
	&&  -\frac{1}{256\pi^2}\left(Y_{\rm u}^{}Y_{\rm u}^\dag+Y_{\rm d}^{}Y_{\rm d}^\dag\right)^{\gamma\delta}\left[Y_\Sigma^{}\frac{\left(3+2L_\Sigma\right)}{M_\Sigma^2}Y_\Sigma^\dag\right]^{\alpha\beta} \;,\qquad
	\label{eq:lQ3w}
	\\
	C^{\alpha\beta\gamma\delta}_{\ell u} &=& -\frac{3}{128\pi^2}\left[Y_\Sigma^{}\frac{\left(3+2L_\Sigma\right)}{M_\Sigma^2}Y_\Sigma^\dag\right]^{\alpha\beta}\left(Y_{\rm u}^\dag Y_{\rm u}^{}\right)^{\gamma\delta} \nonumber \\
	&& +\frac{g_1^2}{576\pi^2}\left[Y_\Sigma^{}\frac{\left(11+6L_\Sigma\right)}{M_\Sigma^2}Y_\Sigma^\dag\right]^{\alpha\beta}\delta^{\gamma\delta} \;,
	\label{eq:lUw}
	\\
	C^{\alpha\beta\gamma\delta}_{\ell d} &=& \frac{3}{128\pi^2}\left[Y_\Sigma^{}\frac{\left(3+2L_\Sigma\right)}{M_\Sigma^2}Y_\Sigma^\dag\right]^{\alpha\beta}\left(Y_{\rm d}^\dag Y_{\rm d}^{}\right)^{\gamma\delta}\nonumber \\
	&& -\frac{g_1^2}{1152\pi^2}\left[Y_\Sigma^{}\frac{\left(11+6L_\Sigma\right)}{M_\Sigma^2}Y_\Sigma^\dag\right]^{\alpha\beta}\delta^{\gamma\delta}  \;,
	\label{eq:ldw}
	\\
	C^{\alpha\beta\gamma\delta}_{\ell e d q} &=& \frac{Y_{\rm d}^{\dag \gamma  \delta}}{32\pi^2}\left\{2Y_l^{\alpha\beta}\tr\left[Y_\Sigma^{}M_\Sigma^{-2}Y_\Sigma^\dag\right]+3\left(Y_\Sigma^{}M_\Sigma^{-2}Y_\Sigma^\dag Y_l^{}\right)^{\alpha\beta}\right\}  \;,
	\label{eq:ledQw}
	\\
	C^{(1)\alpha\beta\gamma\delta}_{\ell e q u} &=& -\frac{Y_{\rm u}^{\gamma\delta}}{32\pi^2}\left\{2Y_l^{\alpha\beta}\tr\left[Y_\Sigma^{} M_\Sigma^{-2}Y_\Sigma^\dag\right]+3\left(Y_\Sigma^{}M_\Sigma^{-2}Y_\Sigma^\dag Y_l^{}\right)^{\alpha\beta}\right\} \;.
	\label{eq:leQUw}
	\end{eqnarray}
	
	\end{itemize}

With all the matching conditions for the Wilson coefficients of the operators up to dim-6, we are able to write down the complete Lagrangian of the SEFT-III at the one-loop level, i.e.,
\begin{eqnarray}
		\mathcal{L}^{}_{\rm SEFT-III} &=& \mathcal{L}^{}_{\rm SM} \left(m^2 \to m^2_{\rm eff}, \lambda\to \lambda^{}_{\rm eff}, Y_l^{}\to Y^{}_{l,{\rm eff}}, Y_{\rm u}^{}\to Y_{\rm u, eff}^{}, Y_{\rm d}^{}\to Y_{\rm d, eff}^{}, g^{}_2 \to g^{}_{2,\rm eff} \right)
		\nonumber
		\\
		&& +  \left[ \left(C^{(5)}_{\rm eff}\right)^{\alpha\beta} O^{(5)}_{\alpha \beta} + {\rm h.c.} \right] +  C^{(1)\alpha \beta}_{Hl \text{-tree}} O^{(1)\alpha \beta}_{Hl} + C^{(3)\alpha \beta}_{Hl \text{-tree}}  O^{(3)\alpha \beta}_{Hl} +  C^{\alpha \beta}_{eH \text{-tree}}  O^{\alpha \beta}_{eH}
		\nonumber
		\\
		 &&  + \sum^{}_i  C^{}_i  O^{}_i \;, \quad
	\label{eq:lagrangian-eft}
\end{eqnarray}
where the original parameter $g\in\{m^2, \lambda, Y^{}_l, Y^{}_{\rm u}, Y^{}_{\rm d}, g^{}_2\}$ in the SM Lagrangian is substituted by its effective counterpart $g_{\rm eff}^{} = g + \delta g^{}_{\rm eff}$, with $\delta g^{}_{\rm eff}$ provided by Eqs.~\eqref{eq:eff-g-1}-\eqref{eq:eff-g-n}. For the dim-5 operator in the second line of Eq.~\eqref{eq:lagrangian-eft}, both the tree-level and one-loop-level coefficients are included, i.e., $C^{(5)}_{\rm eff}=\big(C^{(5)}_{\rm eff}\big)_{\rm tree}+\big(C^{(5)}_{\rm eff}\big)_{\rm loop}$, which are respectively given by Eq.~\eqref{eq:dim5-tree} and Eq.~\eqref{eq:dim5-loop}. In addition,  three dim-6 operators from the tree-level matching are shown separately in the second line and their Wilson coefficients are given in Eq.~\eqref{eq:treematching}. In the last line of Eq.~\eqref{eq:lagrangian-eft}, we formally sum up all the loop-induced dim-6 operators in the Warsaw basis, as explicitly shown in Table~\ref{tab:warsawbasis}, and their coefficients are given in Eqs.~\eqref{eq:3W}-\eqref{eq:leQUw}.

\subsection{Comparison with previous results}
\label{sec:comparison}
The one-loop matching results of the type-III seesaw model should be compared with the previous ones in Ref.~\cite{Du:2022vso}, which are obtained in assumption of the mass degeneracy for three fermionic triplets. To make such a comparison easier, we take the equal-mass limit of our general results in Eq.~\eqref{eq:lagrangian-eft}, i.e., $M_1=M_2=M_3\equiv M$, and find that most of our results are consistent with those in Ref.~\cite{Du:2022vso}. However, we {\it do} observe some mistakes in Ref.~\cite{Du:2022vso}. More explicitly, the matching conditions for the coefficients $\{\delta\lambda, \delta Y^{}_l, C^{}_{eH}, C^{(1)}_{Hl}, C^{(3)}_{Hl}\}$ in the equal-mass limit should be corrected as follows
\bea
\delta\lambda &=& - \operatorname{tr}\left( Y_{\Sigma}^{} Y_{\Sigma}^{\dagger} Y_l^{} Y_l^{\dagger}\right)\left[4(1+L)+\frac{m^2}{2 M^2}(7-2 L)\right]  + \tr\left(Y_{\Sigma}^{} Y_{\Sigma}^{\rm T} Y_{\Sigma}^\ast Y_{\Sigma}^{\dagger}\right)\left[1-\frac{m^2}{6 M^2}(13+6 L)\right] \nonumber\\
&&-\tr\left(Y_{\Sigma}^{\dagger} Y_{\Sigma}^{} Y_{\Sigma}^{\dagger} Y_{\Sigma}^{}\right)\left[5 L+\frac{13 m^2}{3 M^2}\right]  + \frac{m^2}{M^2} \operatorname{tr}\left(Y_{\Sigma}^{\dagger} Y_{\Sigma}^{}\right)\left[4 \lambda+\frac{1}{6} g_2^2(7+2 L)\right] - \frac{2 m^2}{15 M^2} g_2^4 N^{}_{\Sigma} \;, \nonumber \\
\delta Y_l^{pr} &=& \frac{m^2Y_l^{pr}}{M^2}\tr\left(Y_\Sigma^{}Y_\Sigma^\dag\right)+3\left(Y^{}_\Sigma
Y_\Sigma^\dag Y_l^{}\right)^{pr}\left[1+L+\frac{3m^2}{4M^2}(3+2L)\right]  \;,\nonumber \\
C_{eH}^{pr} &=& -\frac{g_2^4}{15M^2} Y_l^{pr} N_\Sigma^{} -\frac{13}{6M^2}\left(Y^{}_l\right)^{pr}\tr\left(Y^{}_\Sigma Y_\Sigma^\dag Y^{}_\Sigma Y_\Sigma^\dag\right) -\frac{1}{12M^2}\left(Y^{}_l\right)^{pr}\tr\left(Y^{}_\Sigma Y_\Sigma^{\rm T} Y_\Sigma^\ast Y_\Sigma^\dag\right)\left(13+6L\right)  \nonumber \\
&&-\frac{1}{4M^2}Y_l^{pr}\tr\left(Y^{}_\Sigma Y_\Sigma^\dag Y^{}_l Y_l^\dag\right)\left(7-2L\right) + \frac{1}{12M^2}Y_l^{pr}\tr\left(Y_\Sigma^\dag Y^{}_\Sigma\right)\left[\left(7+2L\right)g_2^2 + 24 \lambda \right] \nonumber \\
&&+\frac{1}{2M^2}\left(Y^{}_\Sigma Y_\Sigma^\dag Y^{}_l\right)^{pr} \left[\left(41+30L\right) \lambda -2g_1^2 \left(1+3L\right) -4g_2^2 \left(1+3L\right) \right] \nonumber \\
&& -\frac{1}{8M^2}\left(Y^{}_\Sigma Y_\Sigma^\dag Y^{}_\Sigma Y_\Sigma^\dag Y^{}_l\right)^{pr}\left(101+42L\right) +\frac{1}{4M^2}\left(Y^{}_l Y_l^\dag Y^{}_\Sigma Y_\Sigma^\dag Y^{}_l\right)^{pr}\left(17+6L\right)  \nonumber \\
&& +\frac{1}{8M^2}\left(Y^{}_\Sigma Y_\Sigma^\dag Y^{}_l Y_l^\dag Y^{}_l\right)^{pr}\left(7+10L\right) +\frac{7}{8M^2}\left(Y^{}_\Sigma Y_\Sigma^{\rm T} Y_\Sigma^\ast Y_\Sigma^\dag Y^{}_l\right)^{pr}\left(1+2L\right) \nonumber \\
&&-\frac{9}{4M^2} \tr \left(Y^{}_\Sigma Y_\Sigma^\dag\right)\left(Y^{}_\Sigma Y_\Sigma^\dag Y^{}_l\right)^{pr}\left(1+2L\right) \;, \nonumber \\
C^{(1)pr}_{Hl} &=& -\frac{3}{4M^2}\left(Y^{}_\Sigma Y_\Sigma^{\rm T} Y_\Sigma^\ast Y_\Sigma^\dag\right)^{pr} +\frac{3}{16M^2}\left(Y^{}_\Sigma Y_\Sigma^\dag Y^{}_\Sigma Y_\Sigma^\dag\right)^{pr}(17+6L) \nonumber \\
&& + \frac{1}{96M^2}\left[11\left(11+6L\right)g_1^2 - 9\left(7+30L\right)g_2^2\right] \left(Y^{}_\Sigma Y_\Sigma^\dag\right)^{pr} +\frac{g_1^2}{24M^2} \left(5+6L\right) \tr\left(Y^{}_\Sigma Y_\Sigma^\dag\right) \delta^{pr}\nonumber \\
&&+\frac{\left(9+6L\right)}{16M^2}\left(Y^{}_\Sigma Y_\Sigma^\dag Y^{}_l Y_l^\dag + Y^{}_l Y_l^\dag Y^{}_\Sigma Y_\Sigma^\dag\right)^{pr} -\frac{9}{8M^2}\tr\left(Y^{}_\Sigma Y_\Sigma^\dag\right) \left(Y^{}_\Sigma Y_\Sigma^\dag\right)^{pr} \left(1+2L\right)\;,\nonumber \\
C^{(3)pr}_{Hl} &=&-\frac{g_2^4}{15M^2}\delta^{pr}N_\Sigma^{} + \frac{1}{8M^2} \left(Y^{}_\Sigma Y_\Sigma^{\rm T} Y_\Sigma^\ast Y_\Sigma^\dag\right)^{pr} \left(7+6L\right) - \frac{1}{16M^2} \left(Y^{}_\Sigma Y_\Sigma^\dag Y^{}_\Sigma Y_\Sigma^\dag\right)^{pr}(25+6L) \nonumber \\ &&
+\frac{1}{96M^2} \left[3\left(11+6L\right) g_1^2 -\left(127+158L\right) g_2^2 \right] \left(Y^{}_\Sigma Y_\Sigma^\dag\right)^{pr} + \frac{g_2^2}{24M^2}\left(7+2L\right) \tr\left(Y^{}_\Sigma Y_\Sigma^\dag\right)\delta^{pr}
\nonumber \\ &&
+\frac{\left(9+6L\right)}{16M^2}\left(Y^{}_\Sigma Y_\Sigma^\dag Y^{}_l Y_l^\dag + Y^{}_l Y_l^\dag Y^{}_\Sigma Y_\Sigma^\dag\right)^{pr} -\frac{3}{8M^2} \tr\left(Y^{}_\Sigma Y_\Sigma^\dag\right) \left(Y^{}_\Sigma Y_\Sigma^\dag\right)^{pr} \left(1+2L\right)\;,
\label{eq:correctref}
\eea
where $L \equiv \log{\left(\mu^2/M^2\right)}$ has been defined and $N^{}_\Sigma$ denotes the number of fermionic triplets. Two further comments are in order. First, note that our notation for $Y^{}_\Sigma$ is actually the Hermitian conjugation of that in Ref.~\cite{Du:2022vso}, and $m^2 = -\mu_H^2$ is implied. Second, it should be pointed out that the effects of field normalization on $\delta\lambda$ and $\delta Y^{}_l$ haven't been considered in the above equations in order to perform a direct comparison with Ref.~\cite{Du:2022vso}. Additionally, an overall factor of $1/(16\pi^2)$ has been dropped in these coefficients for the same reason.

\section{Further Discussions}
\label{sec:discuss}
\subsection{Radiative decays of charged leptons}
\label{appli}
\begin{figure}[!t]
	\begin{center}
		\includegraphics[width=\linewidth]{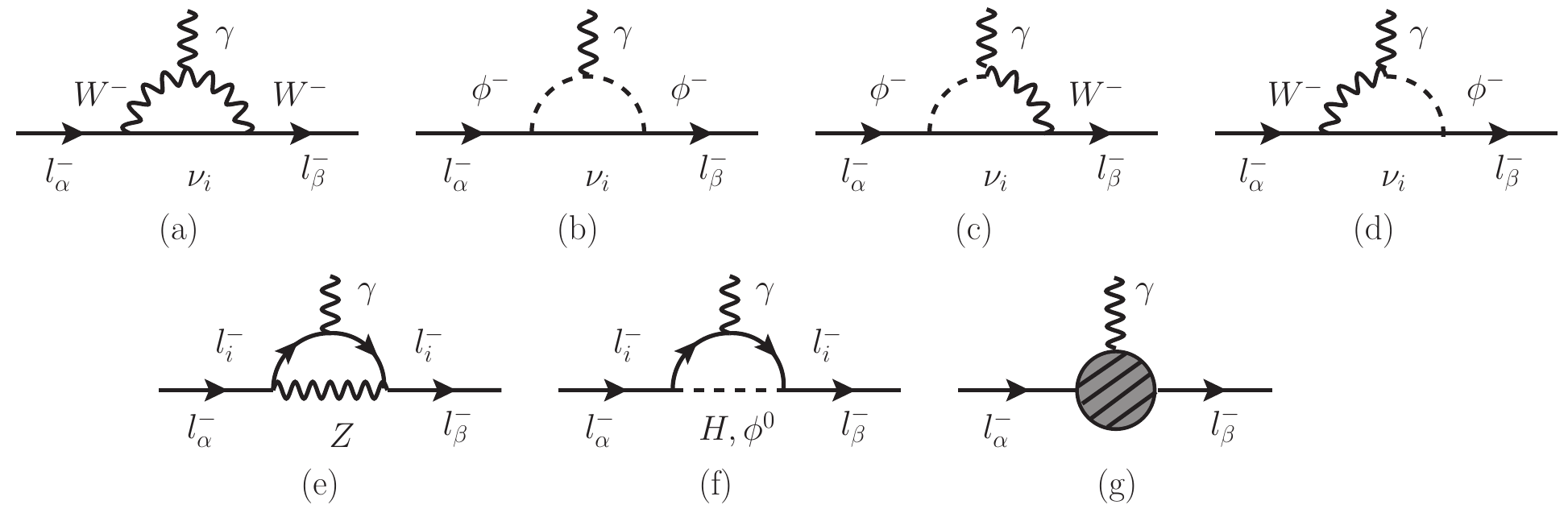}
	\end{center}
	\caption{ The Feynman diagrams for radiative decays of charged leptons $l^-_\alpha \to l^-_\beta + \gamma$ at the one-loop level in the SEFT-III, where the 't Hooft-Feynman gauge has been adopted. Diagrams (a)-(f) are induced by the modified SM interactions after the spontaneous gauge symmetry breaking,  while the diagram (g) is caused by the dim-6 operators $O^{}_{eW}$ and $O^{}_{eB}$ from the one-loop matching in the SEFT-III.
	}
	\label{fig:seft-III}
\end{figure}

The one-loop matching conditions derived in the previous section can be implemented to carry out self-consistent one-loop calculations in the SEFT-III, which are supposed to reproduce the same results in the UV full theory, i.e., the type-III seesaw model. As an explicit example, we calculate the rates of lepton-flavor-violating decays of charged leptons $l^-_\alpha \to l^-_\beta + \gamma$ in the SEFT-III, and compare the results with those in the full type-III seesaw model. In the SEFT-III, after the spontaneous gauge symmetry breaking, the Lagrangian can be written as $\mathcal{L}^{}_\text{ SEFT-III} = \mathcal{L}^{}_{\rm SM}+ \delta \mc{L}$ with $\delta \mc{L}$ being
\begin{eqnarray}
    \delta \mathcal{L}&\supset& -\frac{g_2^{}}{2 c^{}_{\rm w}} \frac{v^2}{2} \overline{l^{}_{\alpha \rm L}} \gamma^\mu l^{}_{\beta {\rm L}} Z^{}_\mu \left(C_{Hl\text{-tree}}^{(1)\alpha\beta} + C_{Hl\text{-tree}}^{(3)\alpha\beta}  \right) - \frac{g_2^{}}{2 c^{}_{\rm w}} \frac{v^2}{2} \overline{\nu^{}_{\alpha \rm L}} \gamma^\mu \nu^{}_{\beta {\rm L}} Z^{}_\mu \left(C_{Hl\text{-tree}}^{(1)\alpha\beta} - C_{Hl\text{-tree}}^{(3)\alpha\beta} \right) \nonumber \\
	&& + \left(\sqrt{2} g^{}_2  \frac{v^2}{2} \overline{l^{}_{\alpha{\rm L}}} \gamma^\mu \nu^{}_{\alpha{\rm L}} W^-_\mu C_{Hl\text{-tree}}^{(3)\alpha\beta} + {\rm h.c.} \right) + \left[\frac{v^2}{2} C_{eH\text{-tree}}^{\alpha \beta} \left( \overline{\ell_{\alpha}} E_{\beta} H \right) + {\rm h.c.}  \right]
    \nonumber \\
    && + \left[ \frac{v}{\sqrt{2}} \left( {c^{}_{\rm w}} C^{\alpha\beta}_{eB} - {s^{}_{\rm w}} C^{\alpha\beta}_{eW} \right) \overline{l^{}_{\alpha \rm L}} \sigma^{}_{\mu\nu} l^{}_{\beta\rm R} F^{\mu\nu} + {\rm h.c.} \right] \nonumber \\
	&=& -\frac{g_2^{}}{2 c^{}_{\rm w}}  \overline{l^{}_{\alpha \rm L}} \gamma^\mu l^{}_{\beta {\rm L}} Z^{}_\mu (RR^\dag )^{}_{\alpha\beta} -\frac{g_2^{}}{2 c^{}_{\rm w}} \overline{\nu^{}_{\alpha \rm L}} \gamma^\mu \nu^{}_{\beta {\rm L}} Z^{}_\mu (RR^\dag )^{}_{\alpha\beta} \nonumber \\
	&&    + \left( \frac{g^{}_2}{2 \sqrt{2}}   \overline{l^{}_{\alpha{\rm L}}} \gamma^\mu \nu^{}_{\alpha{\rm L}} W^-_\mu  (RR^\dag )^{}_{\alpha\beta} + {\rm h.c.} \right) + \left[ \frac{\sqrt{2}}{v} \left( \overline{\ell^{}_{\alpha}} E^{}_{\beta} H \right) (RR^\dag M^{}_l)^{}_{\alpha \beta} + {\rm h.c.}  \right]
    \nonumber \\
	&& - \left[ \frac{e}{32\pi^2 v^2}  \left( \overline{l^{}_{\alpha \rm L}} \sigma^{}_{\mu\nu}  l^{}_{\beta\rm R} F^{\mu\nu}\right) (RR^\dag M^{}_l)^{}_{\alpha \beta} + {\rm h.c.} \right] \; ,
\label{eq:lagrangian-ssb}
\end{eqnarray}
where $v \equiv 2M^{}_W/g^{}_2$, ${c^{}_{\rm w}} \equiv \cos\theta^{}_{\rm w}$, ${s^{}_{\rm w}} \equiv \sin\theta^{}_{\rm w}$, and $R \equiv v Y_\Sigma^{}M_\Sigma^{-1}/\sqrt{2}$ should be noticed. Then, it can be obtained that $RR^\dag = v^2 Y_\Sigma^{} M_\Sigma^{-2} Y_\Sigma^\dag/2$. Furthermore, $\theta^{}_{\rm w} = \arctan(g^{}_1/g^{}_2)$ is the weak mixing angle, and $e=g^{}_1 \cos\theta^{}_{\rm w} = g^{}_2 \sin\theta^{}_{\rm w}$ is the electric charge. Without loss of generality, we work in the mass basis of charged leptons, i.e, $M^{}_l = {\rm diag}\{m^{}_e, m^{}_\mu, m^{}_\tau\}$ with $m^{}_\alpha$ (for $\alpha =e, \mu, \tau$) being the charged-lepton masses. In Eq.~\eqref{eq:lagrangian-ssb}, the tree-level coefficients $C_{Hl\text{-tree}}^{(1)\alpha\beta}$, $C_{Hl\text{-tree}}^{(3)\alpha\beta}$ and $C_{eH\text{-tree}}^{\alpha\beta}$ have been used, as the corresponding vertices appear in the one-loop diagrams in Fig.~\ref{fig:seft-III}. Note that the third line of Eq.~\eqref{eq:lagrangian-ssb} takes account of the contributions from the loop-level operators, i.e., two dim-6 operators $O^{}_{eB}$ and $O^{}_{eW}$, which result in the electromagnetic dipole operator that directly leads to the radiative decays of charged leptons.
With the interactions in Eq.~\eqref{eq:lagrangian-ssb} for the specific process $\mu^-(p^{}_1) \to e^-(p^{}_2) + \gamma(q)$, we can compute the amplitudes for the diagrams in Fig.~\ref{fig:seft-III}
\begin{eqnarray}
	\mathcal{M}_{\text{a-d}}  &=& \frac{e G^{}_{\rm F}}{\sqrt{2}\left(4\pi\right)^2} \left[ U\left(\frac{7}{3}-\frac{\hat{M_\nu^2}}{M_W^2}\right)U^\dag \right]^{}_{e \mu}
	\left[ \overline{u} \left( p^{}_2 \right)  \sigma^{\mu\nu}q^{}_\nu \left( m^{}_e P^{}_{\rm L} + m^{}_\mu P^{}_{\rm R} \right) u\left(p^{}_1\right) \epsilon^\ast_\mu \left(q\right) \right]\;, \nonumber \\
	\mathcal{M}_{\text{e,f}}  &=& \frac{e G^{}_{\rm F}}{3\sqrt{2}\pi^2} \left( c^2_{\rm w}-2 \right)\left( UU^\dag \right)^{}_{e \mu}
	\left[ \overline{u} \left( p^{}_2 \right)  \sigma^{\mu\nu}q^{}_\nu \left( m^{}_e P^{}_{\rm L} + m^{}_\mu P^{}_{\rm R} \right) u\left(p^{}_1\right) \epsilon^\ast_\mu \left(q\right) \right] \;,\nonumber \\
	\mathcal{M}_{\text{g}}  &=&
	\frac{e G^{}_{\rm F}}{\sqrt{2}\left(4\pi\right)^2} \left( 2UU^\dag \right)^{}_{e \mu}
	\left[ \overline{u} \left( p^{}_2 \right)  \sigma^{\mu\nu}q^{}_\nu \left( m^{}_e P^{}_{\rm L} + m^{}_\mu P^{}_{\rm R} \right) u\left(p^{}_1\right) \epsilon^\ast_\mu \left(q\right) \right]\;,
	\label{eq:amp}
\end{eqnarray}
where $U =( \bl{1} + RR^\dag/2)U^{}_0$ is the Pontecorvo-Maki-Nakagawa-Sakata (PMNS) matrix for lepton flavor mixing with $U^{}_0$ being the lowest-order unitary mixing matrix, $G^{}_{\rm F}/\sqrt{2} = g^2_2/(8M^2_W)$ is the Fermi constant, and $\hat{M}^{}_\nu \equiv {\rm diag}\{m^{}_1, m^{}_2, m^{}_3\}$ with $m^{}_i$ (for $i = 1, 2, 3$) being neutrino masses. It is easy to verify that $UU^\dag\simeq \bl{1} + RR^\dag$ and thus the PMNS matrix is non-unitary.\footnote{It is worth noting that the non-unitarity of the PMNS matrix in the type-III seesaw model is induced by the mixing between fermionic triplets and the SM leptons, including both neutrinos and charged leptons~\cite{Abada:2008ea}.} Consequently, the decay rate in the SEFT-III up to $\mc{O}(M_\Sigma^{-2})$ is given by
\begin{eqnarray}
	\Gamma \left( \mu \to e + \gamma \right)
	&\simeq&  \frac{G_{\rm F}^2 e^2 m_\mu^5 }{ 8192 \pi^5 } \left|\left[\frac{13}{3}+ \frac{16}{3}({c^{2}_{\rm w}}-2)
	\right] RR^\dag - \frac{1}{M_W^2}U \hat{M}_\nu^2 U^\dag \right|^2_{e \mu} \;.
	\label{eq:decay-ratio}
\end{eqnarray}
On the other hand, one can also compute the decay rate in the full type-III seesaw model, as has been done in Ref.~\cite{Abada:2008ea}. Based on the calculations in Ref.~\cite{Abada:2008ea}, we expand the results therein with respect to $1/M^{}_\Sigma$ and retain the terms up to $\mc{O}(M_\Sigma^{-2})$. The final result reads
\begin{eqnarray}
	\Gamma\left( \mu \to e + \gamma \right)
	&\simeq& \frac{G_{\rm F}^2 e^2 m_\mu^5 }{ 8192 \pi^5 } \left| \left(\frac{13}{3} + C
	\right) RR^\dag - \frac{1}{M_W^2}U \hat{M}_\nu^2 U^\dag\right|^2_{e \mu} \;.
	\label{eq:ratio-full}
\end{eqnarray}
where the coefficient $C = -6.56$ is given in Ref.~\cite{Abada:2008ea}. Given $s^2_{\rm w} \approx 0.23$, it is straightforward to verify the coefficient $16({c^2_{\rm w}}-2)/3 \approx -6.56$ in Eq.~\eqref{eq:decay-ratio}. As expected, Eq.~\eqref{eq:decay-ratio} agrees perfectly with Eq.~\eqref{eq:ratio-full}. Therefore, starting with the EFT Lagrangian alone, one will be able to carry out complete one-loop calculations of the low-energy observables in a similar way as the simple example shows in this subsection. It is worthwhile to stress that the dim-6 operators $O^{}_{eW}$ and $O^{}_{eB}$ arising from the one-loop matching play an important role~\cite{Zhang:2021tsq}.

\subsection{Beta function of the quartic Higgs coupling}
In the SM, around the energy scale $\mu = \mc{O}(10^{11})$ GeV, the running quartic Higgs coupling $\lambda(\mu)$ becomes negative~\cite{Elias-Miro:2011sqh,Xing:2011aa}, rendering the electroweak vacuum to be unstable. The main reason why $\lambda(\mu)$ declines rapidly with the increasing $\mu$ is that the beta function of $\lambda$ contains a large negative contribution from the top Yukawa coupling $y^{}_t$. In the type-III seesaw model, this problem will be more serious, since $\lambda$ decreases faster than it does in the SM due to the existence of the Yukawa coupling $Y^{}_\Sigma$ of the fermionic triplet. The beta function of $\lambda$ in the type-III seesaw model is~\cite{Goswami:2018jar}:
\begin{equation}
	\begin{aligned}
	\beta(\lambda)\equiv 16 \pi^2 \mu \frac{\mathrm{d} \lambda}{\mathrm{d} \mu} = & \frac{3}{8} g_1^4+\frac{3}{4} g_1^2 g_2^2+\frac{9}{8} g_2^4-3 g_1^2 \lambda-9 g_2^2 \lambda+24 \lambda^2+12 \lambda y_t^2 -6 y_t^4\\
	& +12 \lambda \; \tr\left(Y^{}_{\Sigma} Y_{\Sigma}^{\dagger}\right)-10\; \tr\left(Y^{}_{\Sigma} Y_{\Sigma}^{\dagger} Y^{}_{\Sigma} Y_{\Sigma}^{\dagger}\right) \\
	\equiv& \beta^{\rm SM}_\lambda  +12 \lambda \;  \tr\left(Y^{}_{\Sigma} Y_{\Sigma}^{\dagger}\right)-10\;  \tr\left(Y_{\Sigma} Y_{\Sigma}^{\dagger} Y^{}_{\Sigma} Y_{\Sigma}^{\dagger}\right)\;,
	\end{aligned}
	\label{eq:fullRGE}
\end{equation}
where the beta function $\beta^{\rm SM}_\lambda$ in the SM has been identified in the last line. Since we have matched the type-III seesaw model onto the SMEFT, it is interesting to clarify the relationship between the beta function in the UV full theory and that derived in the EFT. From the EFT perspective, the one-loop renormalization-group (RG) running of $\lambda_{\rm EFT}$ can be studied in the traditional way~\cite{Henning:2014wua}. More explicitly, in the SEFT-III where heavy fermionic triplets have been integrated out, the running of $\lambda_{\rm EFT}$ can be triggered by higher-dimensional operators apart from original SM contributions~\cite{Jenkins:2013zja,Wang:2023bdw}:
\begin{equation}
	\begin{aligned}
		\beta(\lambda_{\rm EFT})\equiv 16 \pi^2 \mu \frac{\mathrm{d} \lambda_{\rm EFT}}{\mathrm{d} \mu}
	=& \beta^{\rm SM}_\lambda +2m^2 \operatorname{tr}\left( C^{}_5 C_5^{\dagger}-\frac{4}{3} g_2^2 C_{H \ell}^{(3)}+4 C_{H \ell}^{(3)} Y^{}_l Y_l^{\dagger} - C_{eH}Y_l^\dag - Y^{}_lC_{eH}^\dag \right)\;,
	\end{aligned}
	\label{eq:EFTRGE}
\end{equation}
where it is sufficient to input the tree-level results of the Wilson coefficients in Eq.~\eqref{eq:treematching} for self-consistent computations at one-loop level. Note that $\lambda_{\rm EFT}\neq \lambda$ here, because the UV theory and the EFT may differ in the UV-divergent behaviors (see, e.g., Ref.~\cite{Penco:2020kvy}, for a review).

However, since we are working on a UV-motivated EFT, we can not only maintain the infrared (IR) information but also even reconstruct the beta functions of the UV theory under certain conditions. In fact, according to the EFT running in Eq.~\eqref{eq:EFTRGE} and one-loop matching result, one can reproduce the running behavior of $\lambda$ in the full type-III seesaw model in Eq.~\eqref{eq:fullRGE}. The key point is that we need to add the contribution from the threshold correction to $\lambda$ into the beta function in Eq.~\eqref{eq:EFTRGE} in the following way
\bea
\beta(\lambda) = \beta(\lambda_{\rm EFT})
+ 16 \pi^2 \mu \frac{\mathrm{d} (\delta\lambda_{\rm eff})}{\mathrm{d} \mu} \;,
\label{eq:rge-eft}
\eea
where $\delta \lambda_{\rm eff}$ is the threshold correction to $\lambda$ given in Eq.~\eqref{eq:eff-g-n} and it depends on the renormalization scale $\mu$. Extracting the $\mu$-dependence of $\delta \lambda_{\rm eff}$ in Eq.~\eqref{eq:eff-g-n} explicitly, one can check that the equality in Eq.~\eqref{eq:rge-eft} indeed holds. With the above demonstration, we emphasize that \emph{the complete expression of $\beta(\lambda)$ in the full theory can be reproduced by summing up the threshold contribution and the beta function $\beta(\lambda^{}_{\rm EFT})$ in the EFT.}

The result in Eq.~\eqref{eq:rge-eft} can be easily understood by focusing on the divergent behaviors from the perspective of the region expansion in the UV theory. First, note that the beta function $\beta(\lambda_{\rm EFT})$ in the EFT is governed by the UV divergence of the soft region of loop momentum in the UV theory. However, this UV divergence can be offset by the IR divergence of the hard region loop momentum, the UV divergence of which corresponds to the true one of the UV full theory. Because of this, the $\mu$-dependence of  $\delta\lambda_{\rm eff}$ in Eq.~\eqref{eq:rge-eft} comes from two sources, i.e., the UV and IR divergences. Consequently, the IR contribution in $16 \pi^2 \mu \mathrm{d}(\delta\lambda_{\rm eff})/\mathrm{d}\mu$ will cancel $\beta(\lambda_{\rm EFT})$ out. In the meanwhile, the remaining part from the UV contribution just produces $\beta(\lambda)$. In this way, Eq.~\eqref{eq:rge-eft} will hold for a general UV theory and its EFT description.

\subsection{Strategy to distinguish among SEFTs}

Thus far all three types of seesaw models have been matched to the SMEFT at the one-loop level. As for the effective operators up to dim-6, these three SEFTs show clear differences between each other, as indicated in Table~\ref{tab:warsawbasis}. Therefore, a natural and interesting question is whether these different dim-6 operators can be implemented to experimentally distinguish among three types of seesaw models. To fully answer this question, one must perform a global-fit analysis of all existing experimental measurements in the framework of the SEFTs and make a model comparison. In this subsection, we just outline a preliminary strategy to look for the answer.

First of all, we notice that the differences among three SEFTs mainly appear as the four-fermion operators. From Table~\ref{tab:warsawbasis}, there are 31 dim-6 operators in the SEFT-I, whereas two additional ones (namely, $O^{}_{W}$ and $O_{qq}^{(3)}$) exist in the SEFT-III. As for the SEFT-II, eight more dim-6 operators are present, including $O_{qq}^{(1)}$, $O^{}_{qe}$ and six four-fermion operators of type ${\rm\overline{R}R\overline{R}R}$, when compared to the SEFT-III. Based on these observations, we propose to search for physical observables that are sensitive to the four-fermion operators in collider experiments. The global-fit analysis of four-fermion operators in the SMEFT indicates that the data from the top-quark sector at the CERN Large Hadron Collider (LHC) may provide very useful information~\cite{Durieux:2022cvf,deBlas:2022ofj}. For instance, the top-pair production is sensitive to four-quark operators, but there occur many degeneracies among these operators and more observables will be helpful to break the degeneracies~\cite{Brivio:2019ius}. Motivated by these studies for the SMEFT, we shall focus on the processes also in the top-quark sector.

Then, we suggest looking into the single-top production at high-energy hadron colliders. As shown in Ref.~\cite{Zhang:2010dr}, the impact of the dim-6 operator $O_{qq}^{(3)}$ on the single-top production is significant. This operator appears in both the SEFT-II and the SEFT-III, but not in the SEFT-I, implying the possibility to distinguish the latter one from the former two.
More explicitly, three operators $O_{qq}^{(3)}$, $O^{}_{uW}$ and $O_{Hq}^{(3)}$ at dim-6 in the SMEFT contribute to the single-top production mainly through the interference with the SM contribution. However, $O^{}_{uW}$ will be neglected in our analysis because it is absent in all three SEFTs. The rest two operators contribute to the production cross-sections of $u + \bar{d} \to t + \bar{b}$ in the $s$-channel and $u + b\to d + t$ in the $t$-channel~\cite{Zhang:2010dr}\footnote{For clarity, we have explicitly factorized out the cutoff-scale dependence $\Lambda^{-2}$ from the Wilson coefficients of dim-6 operators in this subsection such that these coefficients now become dimensionless.}
\begin{equation}
	\begin{aligned}
	\sigma^{}_{u \bar{d} \rightarrow t \bar{b}}= & \left[V_{t b}^2+\frac{2 C_{H q}^{(3)} V^{}_{t b} v^2}{\Lambda^2}\right] \frac{g^4_2 \left(\hat{s}-M_t^2\right)^2\left(2 \hat{s}+M_t^2\right)}{384 \pi  \hat{s}^2\left(\hat{s}-M_W^2\right)^2} +C_{q q}^{(3)} V^{}_{t b} \frac{g^2_2\left(\hat{s}-M_t^2\right)^2\left(2 \hat{s}+M_t^2\right)}{48 \pi \Lambda^2 \hat{s}^2\left(\hat{s}-M_W^2\right)} \;, \\
	\sigma_{u b \rightarrow d t}= & \left[V_{t b}^2+\frac{2 C_{Hq}^{(3)} V^{}_{t b} v^2}{\Lambda^2}\right] \frac{g^4_2\left(\hat{s}-M_t^2\right)^2}{64 \pi \hat{s} M_W^2\left(\hat{s}-M_t^2+M_W^2\right)} -C_{q q}^{(3)} V^{}_{t b} \frac{g^2_2\left(\hat{s}-M_t^2\right) \displaystyle \ln \left[ \frac{\hat{s}-M_t^2+M_W^2}{M_W^2}\right]}{8 \pi \Lambda^2 \hat{s}} \;,
	\end{aligned}
	\label{eq:partonic}
\end{equation}
where $V^{}_{tb}$ denotes the element of Cabibbo-Kobayashi-Maskawa matrix, and only the interference contributions or the terms up to $\mc{O}(\Lambda^{-2})$ are kept, and the tree diagrams with single insertions of $O_{qq}^{(3)}$ and $O_{Hq}^{(3)}$ are considered.
One can see that the contribution of $O_{qq}^{(3)}$ in the $t$-channel is logarithmically enhanced at high energies compared to that of $O_{Hq}^{(3)}$. Such a distinct dependence on the center-of-mass energy $\sqrt{\hat{s}}$ at the parton level signifies the possibility to probe the operator $O_{qq}^{(3)}$. Once the associated Wilson coefficient $C_{qq}^{(3)}$ is found to be nonzero, the SEFT-I can be immediately excluded, since it does not contain this operator.

To estimate the experimental sensitivity to these two relevant Wilson coefficients, we utilize a simple statistical analysis by constructing the $\chi^2$-function as follows
\begin{equation}
	\chi^2 = \frac{\left(\sigma_{\mathrm{exp.}}^s - f\sigma_{\mathrm{EFT}}^s\right)^2}{(\delta\sigma^{}_s)^2} + \frac{\left(\sigma_{\mathrm{exp.}}^t - f\sigma_{\mathrm{EFT}}^t\right)^2}{(\delta\sigma^{}_t)^2} + \frac{(f-1)^2}{(\delta f)^2} \;, \label{eq:chisquare}
\end{equation}
where $\sigma^{s}_{\rm exp.}$ (or $\sigma^{t}_{\rm exp.}$) stands for the measured cross-section in the $s$- (or $t$-) channel at the LHC~\cite{ATLAS:2019hhu} and likewise $\sigma^{s}_{\mathrm{EFT}}$ (or $\sigma^{t}_{\mathrm{EFT}}$) for the expected cross-section in the EFT. While $\delta\sigma^{}_s$ and $\delta\sigma^{}_t$ are the uncertainties of experimental measurements, the theoretical uncertainty of the EFT cross-section has been taken into account by introducing a nuisance parameter $f$ and its error $\delta f$ as a penalty term into the $\chi^2$ function. To obtain the total hadronic cross-section in the proton-proton colliders, one has to convolve the partonic cross-sections in Eq.~\eqref{eq:partonic} with the parton distribution functions (PDF) with the energy threshold being $M^{}_t$. For this purpose, we adopt the \verb|NNPDF31_nlo_as_0118_luxqed| PDFs~\cite{Bertone:2017bme}, and the $s$- and $t$-channel hadronic cross-sections under the $V^{}_{tb}\to 1$ approximation are
\bea
\sigma_{\rm EFT}^t&=&\left(f+\frac{2C_{Hq}^{\left(3\right)}v^2}{\Lambda^2}\right)\sigma_{\rm SM}^t-\frac{C_{qq}^{\left(3\right)}}{\Lambda^2}\left(21.03\ {\rm pb}\right)\cdot\left(1\ {\rm TeV}\right)^2
\nonumber \\
\sigma_{\rm EFT}^s&=&\left(f+\frac{2C_{Hq}^{(3)}v^2}{\Lambda^2}\right)\sigma_{\rm SM}^s+\frac{C_{qq}^{\left(3\right)}}{\Lambda^2}(7.14\ {\rm pb})\cdot\left(1\ {\rm TeV}\right)^2
\eea
where $\sigma_{\rm SM}^s$ and $\sigma_{\rm SM}^t$ are the SM cross-sections for the $s$-channel and the $t$-channel, respectively. From the LHC run-II data at the center-of-mass energy $\sqrt{s}=8$ TeV with the total luminosity $\mc{L}=20.2\ {\rm fb}^{-1}$, the measured single-top cross-sections are $\sigma_{\rm exp.}^s = 4.8^{+1.8}_{-1.5}~{\rm pb}$ and $\sigma_{\rm exp.}^t = 89.6^{+7.1}_{-6.3}~{\rm pb}$, respectively, whereas the NLO+NNLL SM predictions are $5.61 \pm  0.22$ pb and $87.8^{+3.4}_{-1.9}$ pb \cite{Kidonakis:2012rm}. Given these input values, a simplified version of the $\chi^2$-fit analysis can be accomplished by imposing a constraint $\chi^2 - \chi^2_{\rm min} \leq \Delta \chi^2$ at the $68\%$ and $95\%$ confidence levels.
\begin{figure}[!t]
	\begin{center}
		\includegraphics[width=0.45\linewidth]{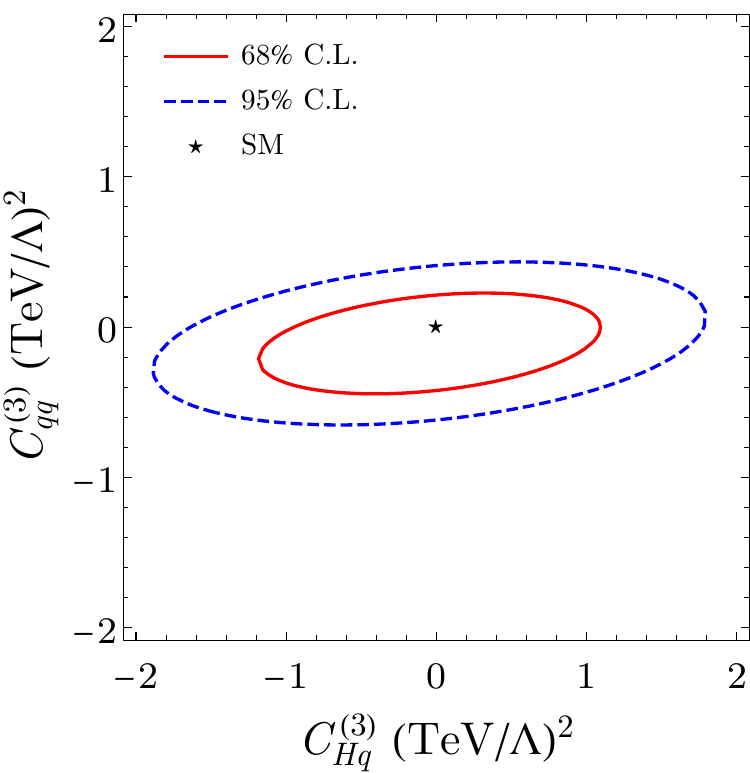}
	\end{center}
	\caption{ The $\chi^2$-fit results for the $s$- and $t$-channel single-top production based on the LHC run-II data, where only two operators $O_{Hq}^{(3)}$ and $O_{qq}^{(3)}$ at the order of ${\cal O}(\Lambda^{-2})$ contribute to the relevant processes. The solid red (or dashed blue) curve indicates the allowed region of two associated Wilson coefficients multiplied by $({\rm TeV}/\Lambda)^2$ at the $68\%$ (or $95\%$) confidence level, while the SM prediction is represented by the black star.
	}
	\label{fig:chi2fit}
\end{figure}

The final results of the $\chi^2$-fit analysis are shown in Fig.~\ref{fig:chi2fit}. At the the $95\%$ confidence level, we obtain the allowed regions $C_{Hq}^{(3)} \cdot ({\rm TeV}/\Lambda)^2 \in [-1.89,1.79]$ and $C_{qq}^{(3)} \cdot ({\rm TeV}/\Lambda)^2 \in [-0.65,0.43]$. Note that the one-loop matching conditions for the Wilson coefficients in the SEFT-II and the SEFT-III give $C_{qq}^{(3)} = -g_2^4/(1920\pi^2)$ and $C_{qq}^{(3)} = -g_2^4/(160\pi^2)$, respectively. Therefore, as the cutoff scale is identified as $\Lambda = M^{}_\Delta$ in the SEFT-II and $\Lambda = M^{}_\Sigma$ in the SEFT-III, one can translate the bounds on the Wilson coefficients in Fig.~\ref{fig:chi2fit} into the lower bounds on the scales of the type-II and type-III seesaw models
\bea
M^{}_\Delta \gtrsim 3.8 \ \text{GeV} \;, \qquad M^{}_\Sigma \gtrsim 13\ \text{GeV}\;.
\eea
It can be seen that these lower bounds on the seesaw scales are rather weak, because the coefficients from one-loop matching are suppressed and the current measurements are not precise enough. In addition,  Fig.~\ref{fig:chi2fit} shows that the LHC run-II data are well consistent with the SM predictions. However, the situation may be greatly improved for future collider experiments. For example, at the HL-LHC, the luminosity will be increased to $3000\ {\rm fb}^{-1}$, such that the experimental uncertainty will be reduced to one percent of the present one. With such a high precision, it is hopefully possible to detect visible deviations from the SM predictions, and the information about the typical scales of type-II and type-III seesaw models can be obtained.

Finally, we point out that further discrimination between the SEFT-II and the SEFT-III is possible but more challenging. When $\sqrt{C}/\Lambda \sim 1~{\rm TeV}^{-1}$ is sizable with $C$ being a general Wilson coefficient of the dim-6 operator, we also need to include the dim-6 squared terms, apart from the interference terms, into the global fit for the single-top production. In this case, there are three additional operators $O_{qq}^{(1)}$, $O^{}_{Hud}$ and $O_{dW}$ that are also involved in the global-fit analysis of the SMEFT\cite{Brivio:2019ius}. Fortunately, the last two operators are absent in the SEFT-II and SEFT-III, as shown in Table~\ref{tab:warsawbasis}, so only $O_{qq}^{(1)}$ is relevant. As this operator happens to be in the SEFT-II, but not in the SEFT-III, it can be utilized to further distinguish between the SEFT-II and the SEFT-III in the measurements of single-top production. Once the dim-6 squared terms are added into the analysis, any indication of nonzero values of $C_{qq}^{(1)}$ will be a smoking-gun signal for the type-II seesaw model, while excluding the type-I and type-III seesaw models. However, a detailed analysis along this line is beyond the scope of this paper and will be left for future works.

\section{Conclusions}
\label{sec:conclude}
In the present paper, we have accomplished the complete one-loop matching of the type-III seesaw model onto the SMEFT via both functional and diagrammatic approaches. The general results for three generations of heavy fermionic triplets are given. A careful comparison with the previous results in Ref.~\cite{Du:2022vso}, where the equal-mass limit has been taken for the fermionic triplets, indicates that some mistakes in the Wilson coefficients in Ref.~\cite{Du:2022vso} need to be corrected. The correct results are also summarized in Sec.~\ref{sec:comparison}.

Furthermore, the low-energy phenomenology of the SEFT-III in three aspects is explored. First, we calculate the rates of lepton-flavor-violating decays of charged leptons in the SEFT-III and demonstrate that the results in the full type-III seesaw model in the large-mass limit can indeed be reproduced when the one-loop matching operators and the associated Wilson coefficients are taken into account. This is a simple example for self-consistent one-loop calculations in the SEFT-III, which will be important to probe neutrino mass models in the precision era of particle physics. Then, we investigate the relationship between the beta function $\beta(\lambda^{}_{\rm EFT})$ of the running quartic Higgs coupling $\lambda^{}_{\rm EFT}$ in the SEFT-III and that $\beta(\lambda)$ in the full type-III seesaw model. It has been shown that $\beta(\lambda)$ in the full theory can be derived by summing up $\beta(\lambda^{}_{\rm EFT})$ in the EFT and the contribution from the $\mu$-dependence of one-loop matching condition for $\lambda$. All these discussions manifest the importance of one-loop matching for phenomenological studies in the EFT. Finally, with the EFTs for three types of seesaw models, we propose a possible way to distinguish among them in collider experiments. For example, the single-top production in the hadron colliders is sensitive to dim-6 four-fermion operators. A novel way to rule out the SEFT-I is to discover the contribution from the operator $O_{qq}^{(3)}$, which is absent in the SEFT-I. The further discrimination between the SEFT-II and the SEFT-III relies on the observation of the dim-6 operator $O_{qq}^{(1)}$, which is more challenging. It is interesting to see whether this strategy can be really implemented in the future analysis of the HL-LHC data to single out the true mechanism for neutrino mass generation.

The origin of neutrino masses definitely calls for new physics beyond the SM. As the canonical seesaw models for neutrino masses usually work at the energy scale far beyond the electroweak scale, precision calculations in their low-energy EFTs will be indispensable when more accurate data are available at the high-energy and high-intensity frontiers. We believe that the one-loop construction of the SEFTs for three types of canonical seesaw models will be very useful for their phenomenological studies at low energies in a systematic way.

\section*{Acknowledgments}
This work was supported by the National Natural Science Foundation of China under grant No.~11835013.


\bibliography{refs}
\bibliographystyle{JHEP}

\end{document}